# Role of intermetallic phases in initiation and propagation of intergranular corrosion of an Al-Li-Cu-Mg alloy

X. Xu[a], M. Hao[b], J. Chen[b], W. He[b], G. Li[b], K. Li[a, d], C. Jiao[c], X.L. Zhong[a], K.L. Moore[a, d], T.L. Burnett[a, e, *], X. Zhou[a, *]

[a] Department of Materials, University of Manchester, Manchester, M13 9PL, UK

[b] Aluminium Alloys and Magnesium Alloys Research Institute, Beijing Institute of Aeronautical Materials, Beijing, 100095, China

[c] Thermo Fisher Scientific, Achtseweg Noord 5, 5651 GG, Eindhoven, the Netherlands.

[d] Photon Science Institute, University of Manchester, Manchester, M13 9PL, UK

[e] Henry Royce Institute for Advanced Materials, The University of Manchester, Manchester, M13 9PL, UK

*Corresponding author:
timothy.burnett@manchester.ac.uk
xiaorong.zhou@manchester.ac.uk

## Abstract

Intermetallic phases in a recently developed Al-Li-Cu-Mg alloy have been investigated to understand their roles in the initiation and propagation processes of intergranular corrosion. Corrosion initiation involves trenching formation in the Al matrix adjacent to the large particles of $Al_7Cu_2$(Fe, Mn) phases. These phases containing Li are electrochemically active and susceptible to self-dissolution via a de-alloying mechanism during corrosion process. The subsurface particles of $Al_7Cu_2$(Fe, Mn) and $Al_{20}Cu_2Mn_3$ phases act as the internal cathodes for continuous corrosion propagation along the particle-matrix interface and the associated grain boundaries. Corrosion propagation along the particle-matrix interface was facilitated by the anodic dissolution of the surrounding Al matrix due to the micro-galvanic interaction with the cathodic intermetallic phases. In addition, $T_1$ ($Al_2CuLi$) precipitates and the isolated particles of $Al_7Cu_2$(Fe, Mn) and $Al_{20}Cu_2Mn_3$ phases were dissolved along the path of corrosion propagation. The dissolved metal ions were redeposited through the network of crevice.

# 1. Introduction

Aluminium-lithium (Al-Li) alloys are widely used in aerospace industry for the applications including fuselage, bulkhead and wing structures [1]. They have excellent specific strength and good fatigue performance that facilitate a higher fuel efficiency and improved performance to modern aircrafts [2]. The modern generation of Al-Li alloys normally contain Cu, Mg, Mn, Zn and Zr with Li in the range of 1 – 2 wt. % [1,2]. They are usually processed using a combination of thermomechanical processing of rolling, solid solution treatment and ageing to produce a microstructure comprising elongated, pancake-shaped grains in combination with constituent intermetallic phases, dispersoids and precipitates [2–5].

The structural components fabricated in Al-Li alloys are susceptible to intergranular corrosion that is associated with localised corrosion attack along grain and sub-grain boundaries. The susceptibility to intergranular corrosion is closely related with a range of factors including alloy composition, compositional heterogeneity at grain boundary and grain stored energy. For instance, the Al-Li-Cu-Mg alloys containing higher contents of Cu and Mg exhibit a higher susceptibility to intergranular corrosion [6,7], whilst the addition of Zn at < 1 wt. % improves the resistance to intergranular corrosion for an Al-1.7Li-2.7Cu -0.3Mg alloy [8]. In addition, intergranular corrosion arising from compositional heterogeneity along the grain boundary area has been extensively reported. For instance, intergranular corrosion has been related with the Cu-depleted precipitate free zones in the vicinity of grain boundary in Al-Li-Cu alloys [9], whilst the continuous distribution of $T_1$ ($Al_2CuLi$) precipitates on grain boundary contributes to a higher susceptibility to intergranular corrosion for a range of Al-Li-Cu alloys [10–13]. Lastly, the grains with relatively high stored energy arising from a higher level of dislocation density are more susceptible to intergranular corrosion as observed in the 2A97 Al-Li-Cu alloy [14–16].

Extensive metallographic and electrochemical investigations further highlight constituent intermetallic phases as an important factor that is associated with intergranular corrosion in a range of Al and Al-Li alloys. For instance, intergranular corrosion preferentially initiates from constituent intermetallic particles in 2090 and 2099 alloys in an aqueous 3.5 wt. % NaCl solution [13,17]. Intergranular corrosion in 2024 alloy also initiates from the clusters of AlCuFeMn intermetallic and S (Al$_2$CuMg) phase particles after immersion in 0.1 M NaCl solution [18–21]. In addition, the α (Al(Fe, Mn, Cu)Si) phases in 6005 alloy are closely related with the initiation and distribution of intergranular corrosion after immersion in an aqueous solution of 10 ml/L HCl and 30g/L NaCl [22,23]. The constituent intermetallic phases that are abundant in Cu and Fe are usually more noble than the surrounding matrix and, in turn, effectively support cathodic reactions in the localised region upon exposure to electrolyte [21,24,25]. This generates a locally basic solution chemistry that undermines the protective oxide layer and causes the dissolution of the surrounding matrix. Micro-galvanic interaction between intermetallic phases and the surrounding matrix further leads to the anodic dissolution of matrix through the most susceptible sites, such as the grain boundary network [19]. The noble intermetallic phases acting as cathodes then support the propagation of intergranular corrosion into the subsurface area through anodic dissolution [20].

Intermetallic phases are electrochemically instable in aqueous electrolyte and actively participate the corrosion process via de-alloying. The de-alloying of intermetallic phase is usually associated with preferential dissolution of less-noble elemental constituents and the formation of Cu-rich remnants with a bi-continuous nanoporous structure [17,21,26–29]. The de-alloying process is self-sustaining through the micro-galvanic interaction between the de-alloyed Cu-rich remnant and the intact part of intermetallic particles, causing the dissolution of active metal components including Al, Mg and Mn [28,29]. It also assists the anodic dissolution of the surrounding matrix by producing an acidic trench environment through the hydrolysis of metal ions [27]. Recently, detailed examinations conducted on the intermetallic particles in 2099 alloy revealed the de-alloying of Li-containing intermetallic phases and the associated intergranular corrosion in the very early stage of immersion in an aqueous 3.5 wt. % NaCl solution [17]. This observation highlights the critical influence of intermetallic phases on initiating intergranular corrosion in Al-Li alloys. However, there is a lack of detailed investigations that thoroughly describe how the intermetallic phases are associated with the initiation and propagation of intergranular corrosion in Al-Li alloys.

Here we investigate the microstructure in a recently developed Al-Li-Cu-Mg alloy both prior to and after immersion in an aqueous 3.5 wt. % NaCl solution. The time-lapse examinations were conducted using multiscale correlative electron and ion microscopy that provides a comprehensive understanding of corrosion morphology and the associated chemical variations including Li distribution. The influence of microstructure on corrosion behaviour is further discussed to provide an in-depth understanding of electrochemical processes associated with the de-alloying of constituent intermetallic phases and intergranular corrosion. The outcomes of current research are intended to provide critical information for improvement of corrosion performance of Al-Li alloys through the optimisation of alloy design and thermomechanical processing.

## 2. Material and experimental procedure

### 2.1 Material

The composition specification of the alloy being investigated is given in Table 1. The as-received material was hot-rolled to a final thickness of 50 mm, followed by solution treatment at ~ 520 °C, then water quenching and pre-stretched to a strain of ~ 4% at room temperature. A two-stage artificial ageing process was then conducted at 120 °C for 20 hours and then 145 °C for 12 hours to reach the peak-aged condition.

**Table 1. The nominal composition (wt. %) of the Al-Li-Cu-Mg alloy investigated in this study.**

| Al | Cu | Li | Mg | Mn |
|---|---|---|---|---|
| bal. | 3.0 – 4.0 | 1.2 – 1.7 | 0.2 – 0.6 | 0.2 – 0.6 |
| Fe | Zr | Si | Zn | |
| ≤ 0.1 | 0.1 – 0.2 | ≤ 0.10 | 0.2 – 0.8 | |

### 2.2 Corrosion immersion testing

Specimens were extracted from the L–LT plane at the mid-thickness (T/2) position for corrosion immersion testing. The specimens measured 20 × 10 mm in length and width and 5 mm in thickness. The specimens were prepared by grinding with SiC grinding papers using water as lubricant, polishing with 3 μm and 1 μm diamond suspensions and chemo-mechanical polishing to remove the subsurface deformed layer that was introduced during the abrasive stages of polishing. The final stage of chemo-mechanical polishing was conducted with great care in a solution of 1:1.5 amorphous colloidal silica suspension (Buehler Mastermet, 0.06 μm) to deionised water to prevent corrosion attack to intermetallic particles and the surrounding matrix,

Figure S1. After polishing, the specimens were first analysed using correlative electron and ion microscopy to investigate the initial microstructure prior to corrosion immersion testing. The specimens were then gently polished in the solution of colloidal silica suspension for ~ 10 s to remove the hydrocarbon contamination deposited in the regions that were analysed using electron and ion microscopy. Further, the specimens were cleaned ultrasonically in an ethanol bath for 20 minutes and dried in a cool air stream prior to immersion testing. Corrosion immersion testing was conducted at room temperature in an aerated solution of 3.5 wt. % NaCl in deionised water. Specimens were immersed for 7 and 30 minutes, followed by thoroughly rinsed in deionised water and dried in a cool air stream for post-mortem examinations of corrosion morphology.

*2.3 Characterisation of initial microstructure*

Orientation mapping of grain structure was performed prior to corrosion immersion testing using electron backscatter diffraction (EBSD). EBSD maps were collected with a step size of 1 μm using an Oxford Instruments Symmetry CMOS acquisition system with the Tescan Mira 3 field emission gun scanning electron microscope (FEG-SEM) at an accelerating voltage of 20 kV and a beam current of 10 nA. Each map covered an area of $2 \times 2$ mm and was composed of ~ 4,000,000 data points collected within 2 hours. The Kikuchi patterns were collected at a pixel resolution of $156 \times 128$ and an exposure time of ~ 1 ms with an indexing rate > 80% confidence. The post-processing of the EBSD data was performed using the HKL Channel 5 software version 5.12.

Constituent intermetallic phases in the same regions that were analysed by EBSD were characterised prior to corrosion immersion testing using the Tescan Mira-3 FEG-SEM system. The chemical composition of these phases was measured using energy dispersive X-ray (EDX) spectroscopy with an Oxford Instruments Xmats 80 SDD detector operated at an accelerating voltage of 10 kV to minimize the interference from the surrounding alloy matrix. EDX chemical analysis was performed automatically at over 500 intermetallic particles using the particle analysis software Oxford Instruments AZtecFeature. The particles > 2 μm in diameter were first identified in the BSE imaging mode using a grey scale thresholding method. The acquisition of EDX data was then conducted locally at the centre of particles at a beam current of 5 nA, a signal count rate of ~ 40,000 cps and a live time of 5 s. The quantitative analysis of EDX data was performed using the Oxford Instruments AZtec software version 5.0 to obtain the average value of chemical composition. The experimentally measured compositions were further compared with the theoretical composition of coarse constituent intermetallics at the thermodynamic equilibrium state. Thermodynamic calculations were carried out using the JMatPro software version 11.2 with the alloy composition as given in Table 1.

High-spatial-resolution analysis of elemental distribution including Li was also conducted prior to immersion using a TOFWERK time-of-flight secondary ion mass spectrometry (ToF-SIMS) analyser added to a Thermo Scientific Helios Tomahawk ion column $Ga^+$ focused ion beam-scanning electron microscope (FIB-SEM). Details of the procedure used for FIB ToF-SIMS analysis was previously reported elsewhere [30]. As a brief introduction here, the analysis was performed using the $Ga^+$ ion beam operated at an accelerating voltage of 30 kV and a beam current at 24 pA over an area 15 × 13 μm in size. The ToF-SIMS signals were collected with a pixel size of ~ 30 nm with 91 frames acquired over 212 s to generate an integrated map with sufficient signal-to-noise ratio. Correlative BSE imaging was conducted in-situ in the same region prior to and after SIMS analysis. BSE micrographs were collected at an accelerating voltage of 5 kV with a pixel size of 14.6 nm using a retractable concentric backscatter (CBS) detector in the Thermo Scientific Helios Nanolab FIB-SEM.

The specimens for transmission electron microscopy (TEM) were prepared by grinding to < 100 μm in thickness, followed by twin jet electropolishing in a solution of 1:3 nitric acid to methanol at a temperature of −25 °C and a potential of 20 V. Precipitate phases were then studied using a Thermo Scientific Talos F200X TEM at 200 kV. Dark-field (DF) TEM micrographs were collected from the $<110>_{Al}$ zone axis to reveal the fine precipitate and dispersoid phases using superlattice reflections corresponding to the δ′ ($Al_3Li$)/θ′ ($Al_2Cu$)/β′ ($Al_3Zr$) and the edge-on $T_1$ precipitates. The misorientation of grain boundaries was also measured using SAEDPs with the procedure as detailed in [31].

*2.4 Post-mortem analysis using correlation electron and ion microscopy*

Backscattered electron (BSE) imaging was conducted prior to and after corrosion immersion testing in the same region that were analysed by EBSD to correlate initial microstructure with the corrosion morphology formed after 7 and 30 minutes of immersion. BSE micrographs were collected using the Tescan Mira 3 FEG-SEM at an accelerating voltage of 5 kV and a beam current of 5 nA. The intermetallic phases abundant in heavy elements have sufficient contrast differential as compared to the surrounding matrix [4]. In total, ~ 100 micrographs were collected at a pixel size of ~ 0.2 μm with each 200 × 200 μm in size over an area measuring 2 × 2 mm. The micrographs were then stitched to generate a montage covering the entire region using the FIJI software version 1.52p. The surface after immersion was further examined in detail using secondary electron (SE) imaging at an accelerating voltage of 5 kV.

Intermetallic particles and the associated corrosion morphology were investigated in the same regions that were analysed by EBSD and BSE imaging using a Thermo Scientific Helios plasma $Xe^+$ FIB-SEM after 7 and 30 minutes of immersion. The use of a $Xe^+$ FIB-SEM is advantageous in reducing the FIB induced artefacts in Al specimens as compared to the $Ga^+$ FIB-SEM systems [32,33]. Automated serial sectioning and image acquisition process were conducted to facilitate the in-situ tomography of the targeted features. In this study, a cross-section was first fabricated in-situ on the targeted particle using a $Xe^+$ ion beam at an accelerating voltage of 30 kV and a beam current of 1.8 nA. SE imaging was then performed at a tilted angle of 52° at an accelerating voltage of the electron beam at 5 kV after each sectioning step. Chemical analysis was performed on the cross-section using EDX with electron beam operated at 10 kV. EDX line profiles were collected at a signal count rate of ~ 20,000 cps and a live time of 1 s per pixel.

High-spatial-resolution chemical analysis was further performed in the subsurface propagation area of intergranular corrosion after 30 minutes of immersion using NanoSIMS 50L. The NanoSIMS chemical analysis was performed correlatively at the same site of intergranular corrosion that was analysed by FIB-SEM tomography on a lamella prepared using the in-situ FIB lift-out technique [34]. The lamella measuring 100 × 50 μm in length and width and 5 μm in thickness was first extracted using the Thermo Scientific Helios plasma $Xe^+$ FIB-SEM at an accelerating voltage of ion beam at 30 kV. The lamella was mounted to a flag post to minimise redeposition during the subsequent stages of surface cleaning at a beam current of 1.8 nA. The NanoSIMS instrument utilises a radio frequency plasma ion source [35] providing an impact energy of 16 keV for O$^-$ primary ions. The smallest primary aperture was used (D1-5, ~ 150 µm diameter) to provide a beam current of 1.1 pA. The secondary ions are separated using a magnetic sector mass analyser with detectors set to collect $^7$Li$^+$, $^{23}$Na$^+$, $^{24}$Mg$^+$, $^{55}$Mn$^+$, $^{56}$Fe$^+$, $^{63}$Cu$^+$ and $^{66}$Zn$^+$ simultaneously. Pre-sputtering of the lamella surface was conducted on with a dose of $1 \times 10^{17}$ ions/cm$^2$ to remove surface contamination and the oxide layer. Chemical maps $15 \times 15$ µm in size were then acquired with a dwell time of 500 µs per pixel with a pixel size of ~ 29 nm over $512 \times 512$ pixels. In total 101 image frames were acquired from the same region and then integrated to generate chemical maps with good signal-to-noise ratio. Data analysis and visualisation (summing image planes, image registration, line scans) was conducted using the FIJI software with the OpenMIMS plugin (Harvard, Cambridge, MA, USA). Correlative BSE imaging was conducted prior to the NanoSIMS analysis in the same region at an accelerating voltage of 5 kV using a CBS detector in the Thermo Scientific Helios Nanolab FIB-SEM.

Lastly, thin-foil specimens were prepared from intergranular corrosion using the Thermo Scientific Helios plasma Xe$^+$ FIB-SEM for examinations using TEM. A lamella measuring $50 \times 20$ µm in length and width was first extracted from the initiation site of intergranular corrosion after 7 minutes of immersion using the in-situ FIB lift-out technique. The lamella was mounted to a flag post and thinned at an accelerating voltage of Xe$^+$ ion beam at 30 kV and beam currents of 1.8 nA then 0.23 nA. The specimen was then cleaned using the ion beam at 5 kV with a beam current of 0.2 nA. The lamella investigated by NanoSIMS after 30 minutes of immersion was also thinned and cleaned using the Xe$^+$ ion beam at the identical conditions. The examination of thin-foil specimens was performed using the Thermo Scientific Talos F200X TEM at 200 kV. Detailed investigation of corrosion morphology was conducted using a combination of methods including high resolution (HR) TEM imaging, scanning transmission electron microscopy (STEM) and EDX.

## 3. Results

*3.1 Initial microstructure prior to corrosion immersion testing*

Figure 1 displays the EBSD grain boundary map and BSE micrographs providing an overview of microstructure together with a BSE micrograph and the correlative FIB

ToF-SIMS chemical maps showing details of constituent intermetallic phases and elemental distribution. Table 2 compares the experimentally measured compositions of constituent intermetallic phases with the theoretical compositions at the thermodynamic equilibrium state as calculated using JMatPro software.

Figure 1a shows that the initial microstructure prior to corrosion immersion testing is predominantly composed of pancake-shaped unrecrystallised grains in combination with a small fraction of recrystallised grains (i.e. area fraction ~ 10%) without internal substructure. Figure 1b further reveals the coarse constituent intermetallic phases that are preferentially distributed on high angle grain boundaries. BSE imaging at a higher magnification reveals the particles showing different contrast differentials, Figure 1c. The comparison as illustrated in Table 2 further shows a good agreement for the contents of Al, Cu and Mn between the experimentally measured composition and the compositions of equilibrium phases predicted using thermodynamic calculation. This confirms that the intermetallic particles with a shiny bright appearance in Figure 1c are the $Al_7Cu_2(Fe, Mn)$ phases, whilst the ones appear in grey are the $Al_{20}Cu_2Mn_3$ phases. Zn and Mg were also detected in the constituent intermetallic particles at a trace level with no significant difference between the $Al_7Cu_2(Fe, Mn)$ and $Al_{20}Cu_2Mn_3$ phases.

BSE imaging and the correlative high-spatial-resolution FIB ToF-SIMS analysis further reveal the Li-rich needle-shaped $T_1$ phase particles, Figures 1d and 1e. These quench-induced $T_1$ precipitates that are ~ 5 μm in length were formed during solid solution treatment as revealed in a previous study conducted on the same material [36]. The $Al_7Cu_2(Fe, Mn)$ particle also have a similar signal intensity of $^7Li^+$ as compared to the surrounding matrix, whilst the $^7Li^+$ signal from the $Al_{20}Cu_2Mn_3$ particles is considerably lower. In addition, the $^{24}Mg^+$ signal was detected from the $Al_7Cu_2(Fe, Mn)$ particle, whilst no $^{24}Mg^+$ signal was obtained from the $Al_{20}Cu_2Mn_3$ particles. The FIB ToF-SIMS maps further reveal the segregation of Li and Mg on grain boundaries and the surface of intermetallic particles, Figures 1e and 1f. As previously reported, this is due to the presence of continuous trace of $T_1$ precipitates that were formed during artificial ageing [30]. Detailed TEM examinations further reveal the PFZs close to the quench-induced $T_1$ precipitates are ~ 100 nm in width, whilst the PFZs adjacent to the ageing-induced intergranular $T_1$ precipitates are ~ 30 nm in width, Figure S2. The δ′/θ′ precipitates measuring 5 – 30 nm, the needle-shaped intragranular $T_1$ precipitates < 30 nm in length and the circular $\beta'$ dispersoids < 50 nm in diameter were also observed using DF-TEM imaging with the corresponding reflections, Figure S2.

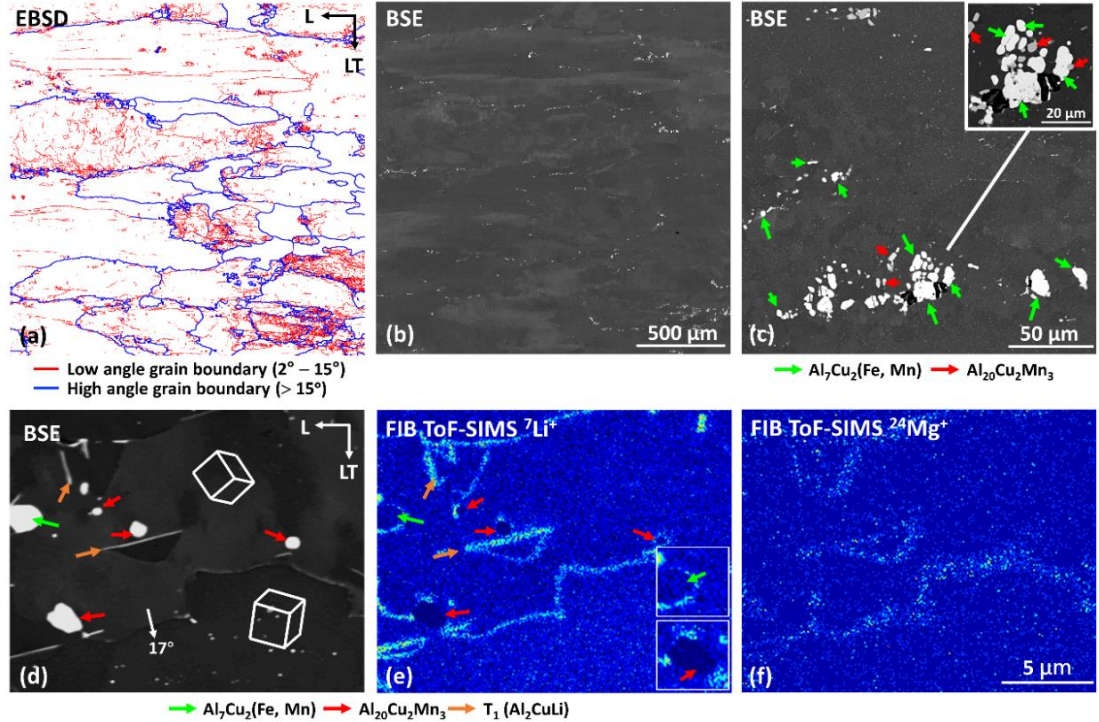

**Figure 1. (a) EBSD grain boundary map and (b) the correlative BSE micrograph providing an overview of the initial microstructure prior to corrosion immersion testing. Low angle and high angle grain boundaries are indicated in red and blue, respectively. (c) A BSE micrograph at a higher magnification showing details of constituent intermetallic phases. The $Al_7Cu_2(Fe, Mn)$ and $Al_{20}Cu_2Mn_3$ phases were identified using EDX as indicated by green and red arrows, respectively. (d) A BSE micrograph and (e, f) the correlative FIB ToF-SIMS chemical maps showing the distribution of $^7Li^+$ and $^{24}Mg^+$. The $Al_7Cu_2(Fe, Mn)$, $Al_{20}Cu_2Mn_3$ and quench-induced $T_1$ ($Al_2CuLi$) phases were identified based on particle morphology and EDX as indicated by green, red and orange arrows, respectively. The insets and arrow in (d) indicate grain orientation and grain boundary misorientation as measured by EBSD.**

**Table 2. A comparison between the experimentally measured composition of constituent intermetallic phases and the theoretical compositions at the thermodynamic equilibrium state.**

| Thermodynamic calculation | | | | | Experimental measurement* | | | | |
|---|---|---|---|---|---|---|---|---|---|
| | $Al_7Cu_2(Fe, Mn)$ | | $Al_{20}Cu_2Mn_3$ | | | $Al_7Cu_2(Fe, Mn)$ | | $Al_{20}Cu_2Mn_3$ | |
| Element | wt. % | at. % | wt. % | at. % | Element | wt. % | at. % | wt. % | at. % |
| Al | 50.8 | 70.0 | 64.9 | 80.0 | Al | 53.4 ± 2.2 | 72.2 ± 2.1 | 61.6 ± 1.7 | 77.7 ± 1.2 |
| Mn | 0.0 | 0.0 | 19.8 | 12.0 | Mn | 3.3 ± 1.3 | 2.2 ± 0.9 | 18.0 ± 0.9 | 11.1 ± 0.6 |
| Fe | 15.0 | 10.0 | 0.0 | 0.0 | Fe | 8.8 ± 2.7 | 5.8 ± 1.8 | 2.8 ± 0.7 | 1.7 ± 0.4 |
| Cu | 34.2 | 20.0 | 15.3 | 8.0 | Cu | 33.8 ± 1.7 | 19.3 ± 1.2 | 17.2 ± 1.1 | 9.1 ± 0.1 |

*Zn, Mg and Si were experimentally detected at a trace level < 0.5 wt. %.

*3.2 Post-mortem analysis after 7 minutes of immersion*

Figure 2 shows the correlative BSE micrographs and EBSD maps comparing the surface of specimen prior to and after 7 minutes of immersion with the BSE micrographs showing details of intermetallic particles.

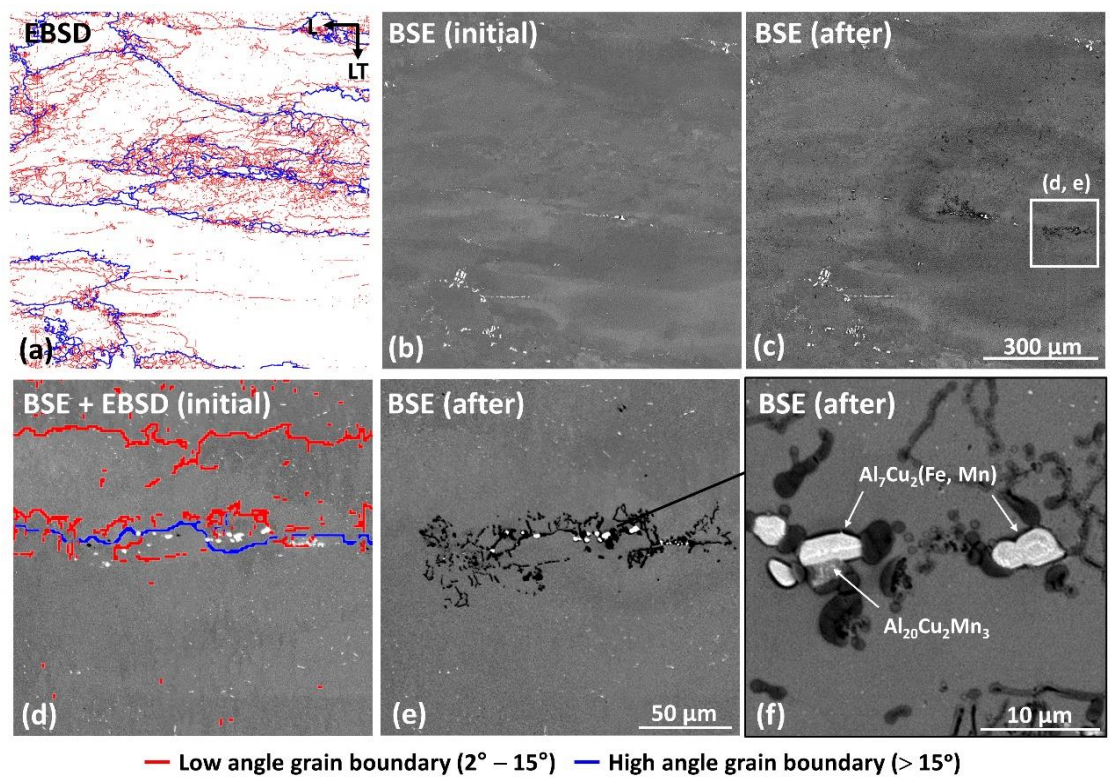

**Figure 2. (a) EBSD grain boundary map and (b, c) the correlative BSE micrographs comparing the surface of specimen prior to and after 7 minutes of immersion. (d) A BSE micrograph overlaid with EBSD grain boundary map detailing the surface prior to immersion and (e) the correlative BSE micrograph showing intergranular corrosion morphology. (f) A BSE micrograph showing details of the associated intermetallic particles as identified using EDX.**

The comparison of sample surface as illustrated in Figures 2a – 2c shows the presence of intergranular corrosion associated with two clusters of intermetallic phases associated with a dense network of grain boundary in the early stage of corrosion immersion testing after 7 minutes of immersion. A ring of corrosion product with a diameter of > 500 μm was also observed in the surrounding area. Figures 2d and 2e reveal a strong correlation of intergranular corrosion with the low angle and high angle grain boundaries associated with intermetallic phases. Figure 2f further reveals trenches measuring 100 – 200 nm in width in the matrix adjacent to the $Al_7Cu_2$(Fe, Mn) particles. In addition, corrosion morphology was observed on the surfaces of both $Al_7Cu_2(Fe, Mn)$ and $Al_{20}Cu_2Mn_3$ phases.

Figure 3 shows the SE micrographs collected from cross-section at the site of intergranular corrosion showing an overview of corrosion morphology and details of the associated intermetallic particles.

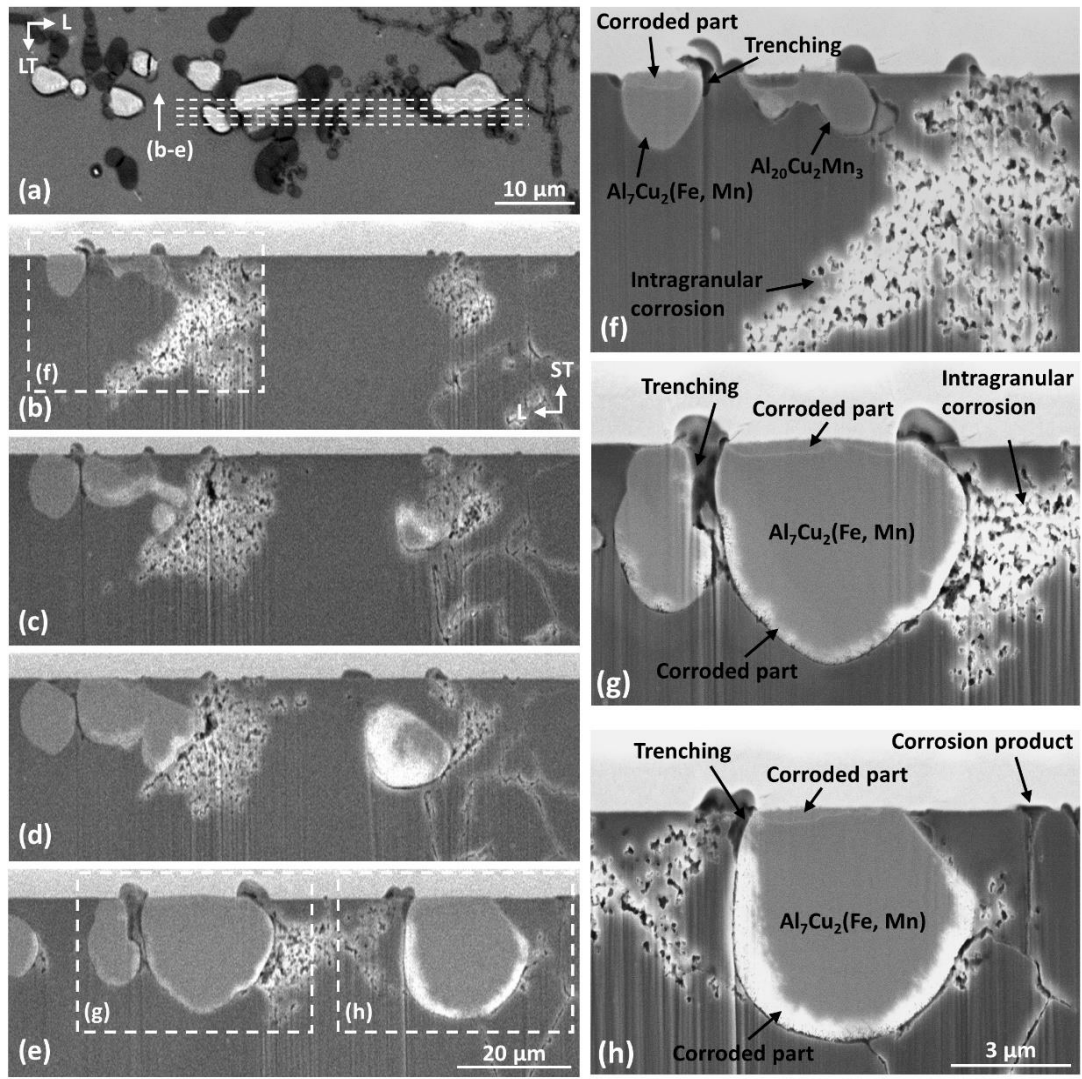

**Figure 3. SE micrographs showing (a) surface and (b-e) cross-sections at the site of intergranular corrosion after 7 minutes of immersion. The cross-sections were fabricated using a FIB-SEM at the locations indicated by dash lines in (a). (f-h) SE micrographs showing details of the associated intermetallic particles from the locations as indicated in (c, e).**

Figures 3a – 3e clearly reveal intergranular corrosion morphology and the associated intermetallic particles that were identified to be the $Al_7Cu_2(Fe, Mn)$ and $Al_{20}Cu_2Mn_3$ phases using EDX. The subsurface attack of intergranular corrosion preferentially propagates in the L direction with a maximum depth of corrosion attack in the ST

direction at ~ 20 μm, Figure S3. Detailed SE micrographs from the intermetallic particles further confirm the presence of trenches in the matrix adjacent to the $Al_7Cu_2$(Fe, Mn) phases, whilst trenches were not observed adjacent to the $Al_{20}Cu_2Mn_3$ phases, Figures 3f – 3h. In addition, different characteristics of corrosion morphology were observed in the intermetallic phases at top surface and from ~ 1 μm beneath the top surface along the interface with the surrounding matrix. The corroded areas are < 200 nm and 200 – 500 nm in width at top surface and close to the subsurface particle-matrix interface, respectively. Corrosion products are also accumulated at top surface adjacent to the intermetallic particles or connected with grain boundaries. Intergranular corrosion attack leads to subsurface crevices at the particle-matrix interface and grain boundaries with a width of ~ 50 nm. Crystallographic corrosion attack within the grain/sub-grain interiors was also observed adjacent to the intermetallic particles.

Detailed examinations were further conducted using TEM with a particular focus on the initiation site of intergranular corrosion. Figure 4 shows a HAADF-STEM and the correlative STEM-EDX maps from the intermetallic particle as presented in Figure 3h, with the STEM-EDX line-profiles detailing the chemical variations across top surface and the interface with the surrounding matrix.

Figures 4a – 4g clearly shows the selective depletion of Al, Mg and Mn in the corroded part of $Al_7Cu_2$(Fe, Mn) phase with relative abundance of O in the same regions. Figures 4h and 4j further reveal nano-porous structure close to top surface and the subsurface particle-matrix interface of intermetallic particle. The porous structure is relatively coarser close to the subsurface particle-matrix interface. The STEM-EDX line profile across top surface shows that the contents of Al, Mg, Si and Mn are considerably lower in the de-alloyed region (i.e. Region i-I) than the intact part of intermetallic particle (i.e. Region i-II), while the contents of Cu, Fe and Zn are slightly decreased in the de-alloyed part, Figure 4i. Figure 4k reveals that the contents of Al, Mg, Si, Fe and Mn are evidently lower towards the particle-matrix interface, whilst the relative enrichment of Cu and Zn was observed in the region < 200 nm away from the interface (i.e. Region k-II). In addition, a thin layer abundant in Al and O was observed with a thickness measuring ~ 50 nm in the matrix adjacent to the de-alloyed particle (i.e. Region k-III).

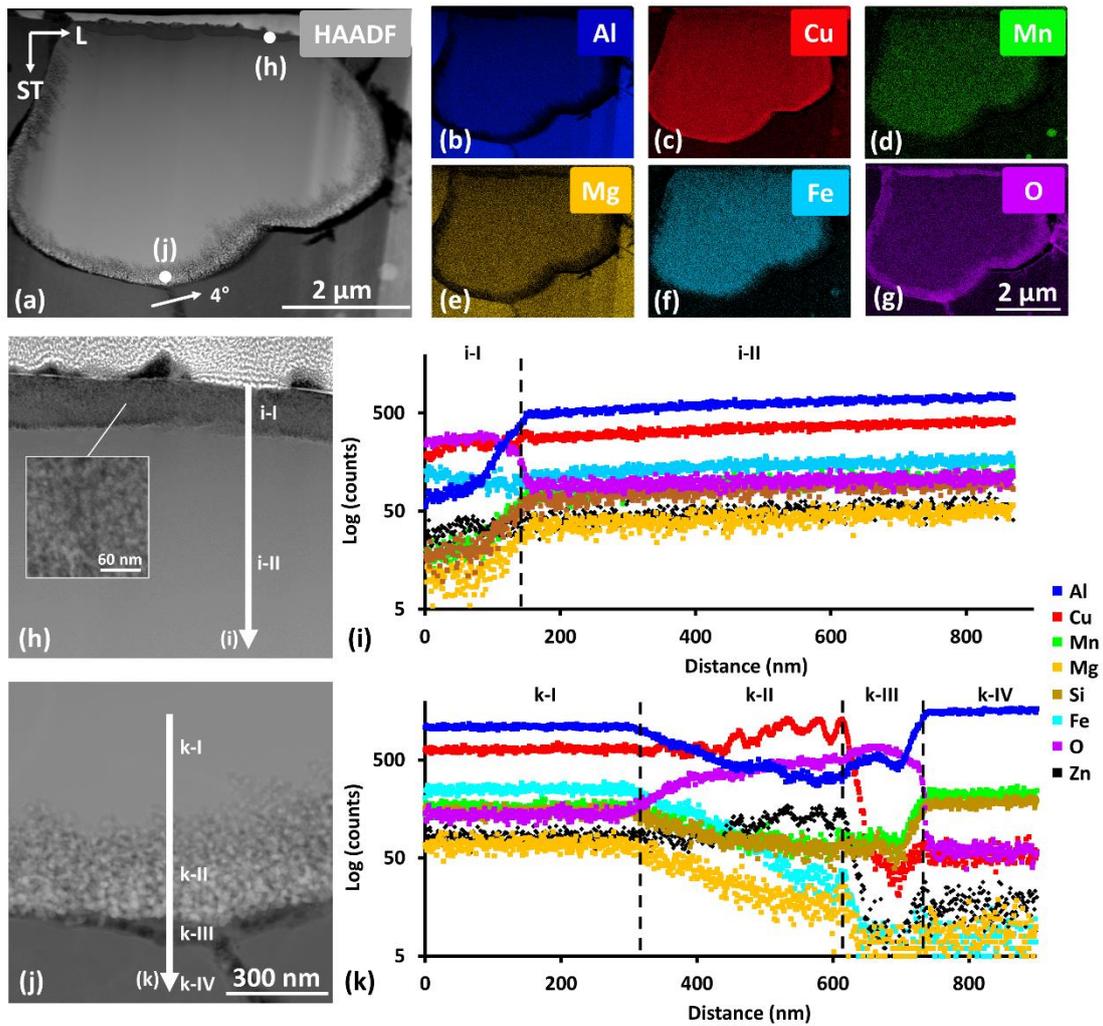

**Figure 4. (a) A HAADF-STEM micrograph and (b-g) STEM-EDX maps providing an overview of an Al$_7$Cu$_2$(Fe, Mn) intermetallic particle at the initiation site of intergranular corrosion after 7 minutes of immersion. The number in (a) indicates the misorientation of grain boundary. (h, j) The HAADF-STEM micrographs at a higher magnification showing details of corrosion morphology at the locations as indicated in (a), with (i, k) STEM-EDX line profiles with a thickness of 50 pixels showing chemical variations across top surface and the particle-matrix interface. The inset in (h) shows the detail of corrosion morphology close to top surface.**

Figure 5 shows the HR-TEM micrographs collected from top surface and the subsurface particle-matrix interface from the Al$_7$Cu$_2$(Fe, Mn) particle. Figures 5a and 5b reveal a nanocrystalline structure containing Cu$_2$O and Cu in the corroded region close to top surface. The reflections corresponding to Cu, Cu$_2$O and Cu$_3$O$_4$ were identified from the corroded structure close to the subsurface interface, Figures 5c and 5d.

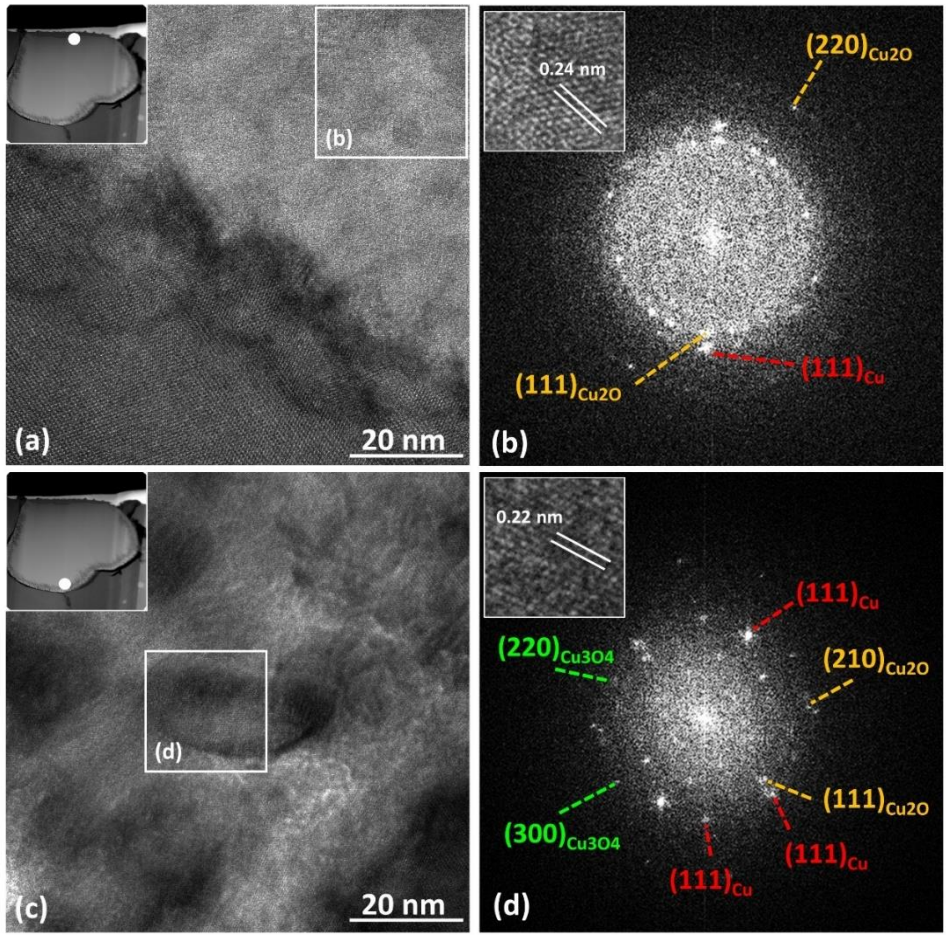

**Figure 5.** HR-TEM micrographs collected from (a) top surface and (c) the subsurface particle-matrix interface with (b, d) fast Fourier transformation (FFT) patterns from the boxed regions. The insets in (a) and (c) indicate the locations where the HR-TEM micrographs were collected. The insets in (b) and (d) show details in the targeted regions.

*3.3 Post-mortem analysis after 30 minutes of immersion*

Figure 6 shows the correlative BSE micrographs and EBSD maps comparing the surface of specimen prior to and after 30 minutes of immersion with the BSE micrographs showing details of intermetallic particles.

The correlative comparison as illustrated in Figures 6a – 6c reveals intergranular corrosion associated with a cluster of intermetallic phases after 30 minutes of immersion. The presence of corrosion products was also observed adjacent to the clusters of intermetallic phases that are not related with intergranular corrosion. Figures 6d and 6e further reveal intergranular corrosion attack through the network of low angle and high angle grain boundaries. The intermetallic particles associated with intergranular corrosion are predominantly the $Al_7Cu_2$(Fe, Mn) phases as identified using EDX, Figure 6f. The accumulation of corrosion products adjacent to these intermetallic phases is not as evident as compared to the intermetallic particles that are not related with intergranular corrosion, Figure S4.

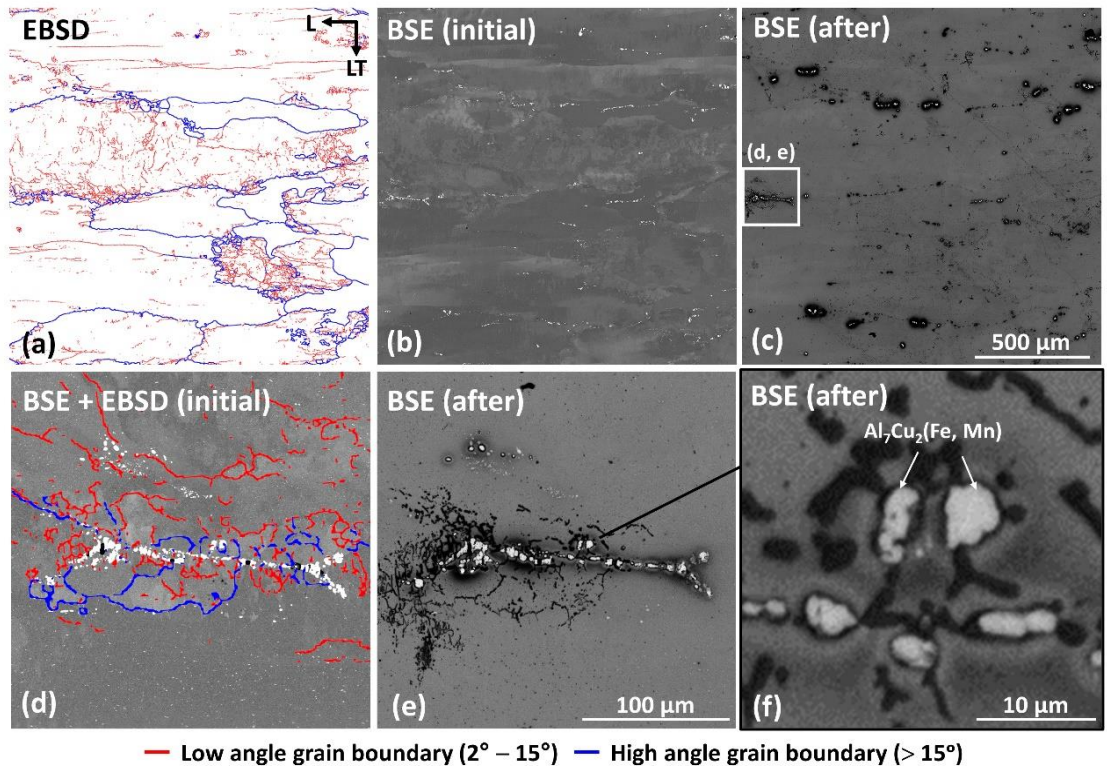

**Figure 6. (a) EBSD grain boundary map and (b, c) the correlative BSE micrographs comparing the surface of specimen prior to and after 30 minutes of immersion. (d) A BSE micrograph overlaid with EBSD grain boundary map detailing the surface prior to immersion and (e) the correlative BSE micrograph showing intergranular corrosion morphology. (f) A BSE micrograph showing details of the associated intermetallic particles as identified using EDX.**

Figure 7 shows the SE micrographs collected from the cross-section of intergranular corrosion after 30 minutes of immersion. Figures 7a – 7d reveal corrosion morphology predominantly with intergranular corrosion and crystallographic corrosion within the grain interiors. The maximum depth of corrosion attack after 30 minutes of immersion is ~ 20 μm in the ST direction, Figure S5. The intermetallic particles associated with intergranular corrosion are isolated from the adjacent matrix due to corrosion attack along the particle-matrix interface. These phases are clusters of the $Al_7Cu_2$(Fe, Mn) and $Al_{20}Cu_2Mn_3$ phases as identified using EDX. Figure 7e – 7h further reveal the intermetallic phases associated with intergranular corrosion in the subsurface region. The corrosion attack preferentially propagates through the particle-matrix interface, then follow along the network of the associated grain boundaries.

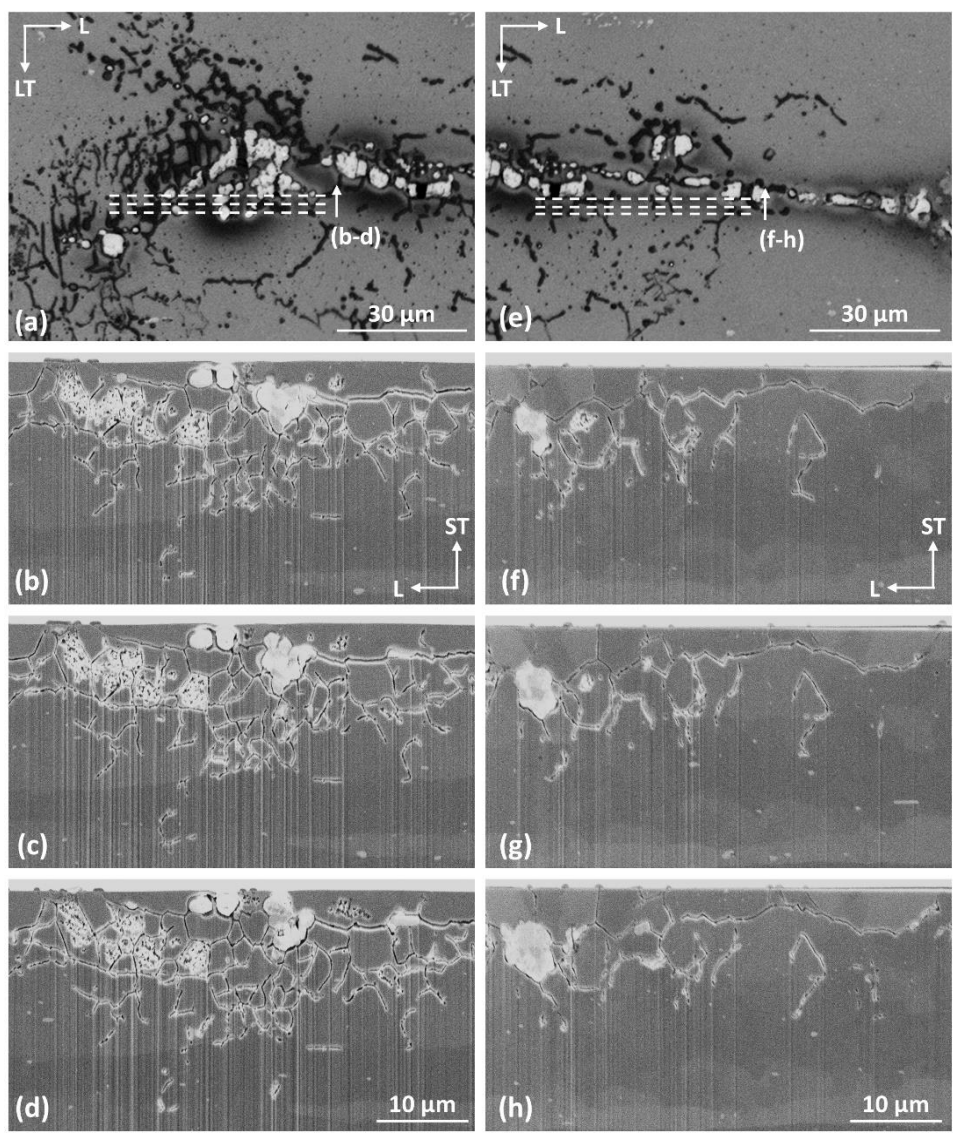

**Figure 7. SE micrographs showing (a, e) surface and (b-d, f-h) cross-sections at the site of intergranular corrosion after 30 minutes of immersion. The cross-sections were fabricated using a FIB-SEM at the locations indicated by dash lines in (a) and (e).**

Detailed examinations were further performed in the subsurface propagation area of intergranular corrosion using NanoSIMS and TEM on a lamella extracted using a FIB-SEM, Figure S6. Figure 8 first shows the NanoSIMS chemical maps and the correlative BSE micrographs obtained from the cross-section of intergranular corrosion after 30 minutes of immersion. The chemical maps clearly reveal the depletion of Li and Mg in the intermetallic phases associated with intergranular corrosion, whilst these particles are relative abundant in Mn and Cu. The 'spots' showing strong intensity of $^7Li^+$ signal was also observed within or adjacent to the crevice of intergranular corrosion along the corroded grain boundary, Figure 8b. These 'spots' are related with the locations showing a strong intensity of $^{23}Na^+$ signal, Figure 8c. In addition, strong intensities of $^{55}Mn^+$ and $^{63}Cu^+$ were collected along the grain boundary, Figures 8e and 8f.

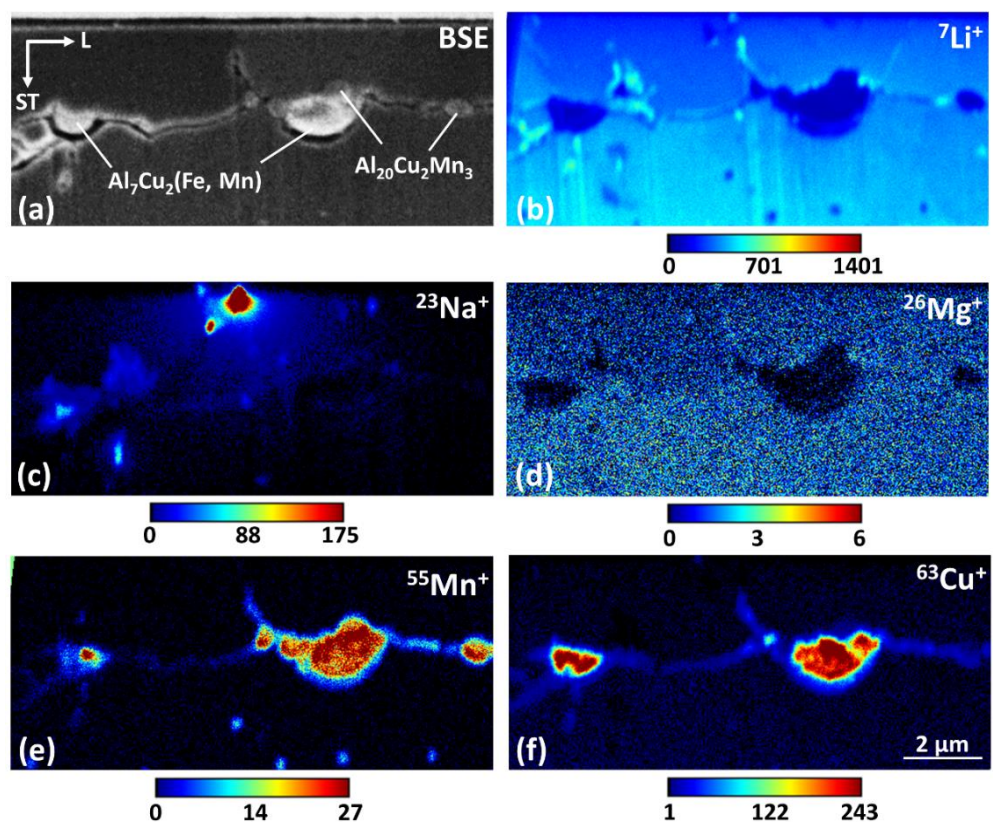

**Figure 8. (a) A BSE micrograph and the correlative NanoSIMS chemical maps showing the distribution of (b) $^7Li^+$, (c) $^{23}Na^+$, (d) $^{26}Mg^+$, (e) $^{55}Mn^+$ and (f) $^{63}Cu^+$. The intermetallic phases as indicated in (a) were identified using EDX. The location where the NanoSIMS analysis was performed is indicated in Figure S6.**

Figure 9 displays a HAADF-STEM micrograph and the correlative STEM-EDX chemical maps showing details of intergranular corrosion and the associated intermetallic phases in the region as indicated in Figure S6. Table 3 further shows the chemical composition for the remnants of intermetallic phases based on quantitative EDX analysis.

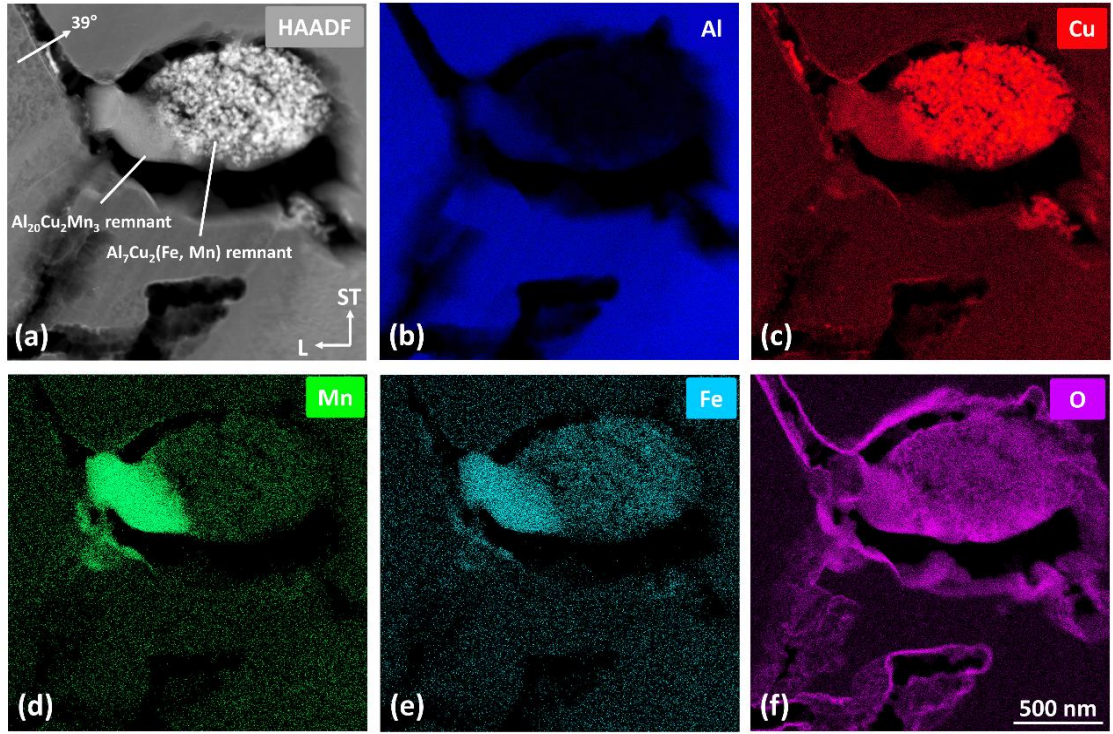

**Figure 9.** (a) A HAADF-STEM micrograph and (b-f) the correlative STEM-EDX chemical maps showing details of the remnants of $Al_7Cu_2$(Fe, Mn) and $Al_{20}Cu_2Mn_3$ phases in the subsurface propagation area of intergranular corrosion after 30 minutes of immersion. The number in (a) indicates the misorientation of grain boundary. The location where the analysis was performed is indicated in Figure S6.

**Table 3.** Experimentally measured composition for the remnants of $Al_7Cu_2$(Fe, Mn) and $Al_{20}Cu_2Mn_3$ phases based on quantitative STEM-EDX analysis.

|  | $Al_7Cu_2$(Fe, Mn) | | $Al_{20}Cu_2Mn_3$ | |
|---|---|---|---|---|
| Element | wt. % | at. % | wt. % | at. % |
| Al | 10.1 | 6.2 | 50.7 | 44.7 |
| Mn | 0.1 | 0.2 | 8.7 | 15.6 |
| Fe | 0.3 | 0.4 | 1.2 | 2.1 |
| Cu | 55.5 | 80.2 | 10.7 | 22.1 |
| O | 32.0 | 11.7 | 27.6 | 14.4 |

*Mg, Zn and Si were detected at a trace level < 1 wt. %.

Figure 10 further shows the HAADF-STEM micrographs collected at a higher magnification showing details of corrosion morphology along corroded grain boundary and the associated intermetallic phases. A STEM-EDX line profile is provided to detail the chemical variation across the internal surface of corroded grain boundary. An FFT pattern is also provided to exhibit crystallography for the remnant of a $Al_7Cu_2$(Fe, Mn) particle.

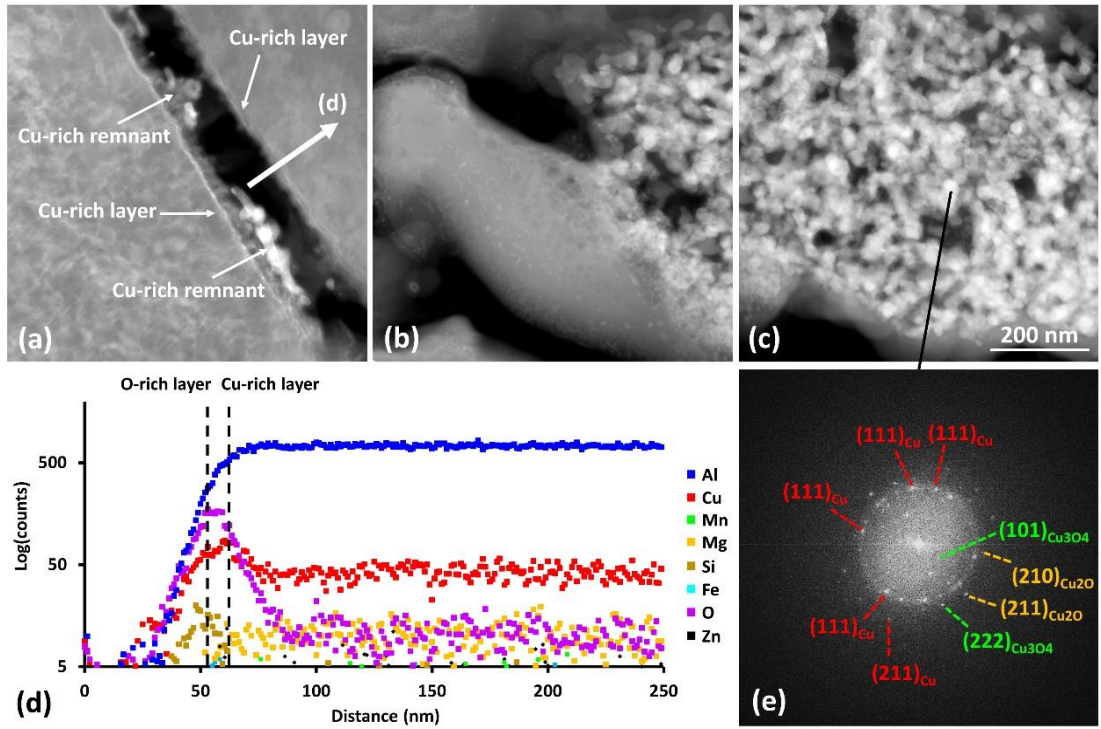

**Figure 10.** HAADF-STEM micrographs showing details of (a) corroded grain boundary and the corroded remnants of (b) $Al_{20}Cu_2Mn_3$ and (c) $Al_7Cu_2(Fe, Mn)$ phases. (d) A STEM-EDX line profile with a line thickness of 50 pixels showing chemical variation across the internal surface of corroded grain boundary in (a). (e) An FFT pattern of a HR-TEM micrograph collected from the remnant of $Al_7Cu_2(Fe, Mn)$ phase as shown in (c).

HAADF-STEM imaging reveals the nanoporous structure for the remnant of a corroded $Al_7Cu_2$(Fe, Mn) intermetallic phase with the adjacent $Al_{20}Cu_2Mn_3$ phase showing no porous structure, Figures 9a, 10b and 10c. The corrosion of $Al_7Cu_2$(Fe, Mn) phase leads to a Cu, O-rich remnant with relative depletion of Al, Fe and Mn as revealed by STEM-EDX analysis, Figures 9b – 9f and Table 3. HR-TEM imaging further identifies the reflections corresponding to Cu, $Cu_2O$ and $Cu_3O_4$ from the Cu, O-rich remnant, Figure 10e. Table 3 also shows that the remnant of $Al_{20}Cu_2Mn_3$ phase contains a higher content of O and lower contents of Al, Mn, Fe and Cu than the original composition as indicated in Table 2.

In the same region, intergranular corrosion propagates along a high angle grain boundary (i.e. ~ 39° in misorientation) associated with the cluster of intermetallic phases, causing a crevice measuring ~ 120 nm in width, Figure 9a. Corrosion attack at the particle-matrix interface also leads to crevices measuring ~ 100 nm and ~ 270 nm in width as observed above and beneath the intermetallic particles, respectively. The internal surface of crevice is enriched in Cu and O, while the enrichment of Fe and Mn was also observed within the crevice close to the $Al_{20}Cu_2Mn_3$ particle, Figures 9b – 9f. STEM-EDX line profile further reveals a thin layer abundant in Si and O at the internal surface of corroded grain boundary, whilst the Cu-rich layer is in the matrix adjacent to the crevice, Figure 10d. In addition, nanoscale particles of corroded remnants measuring < 200 nm in length were observed within the crevice of corroded grain boundary adjacent to the intermetallic phases, Figures 9a and 10a. STEM-EDX analysis reveals the enrichment of Cu and O for these remnants, Figures 9c and 9f.

Figure 11 shows a HAADF-STEM micrograph and the correlative STEM-EDX chemical maps showing details of intergranular corrosion and the associated quench-induced $T_1$ precipitate in the region as indicated in Figure S6.

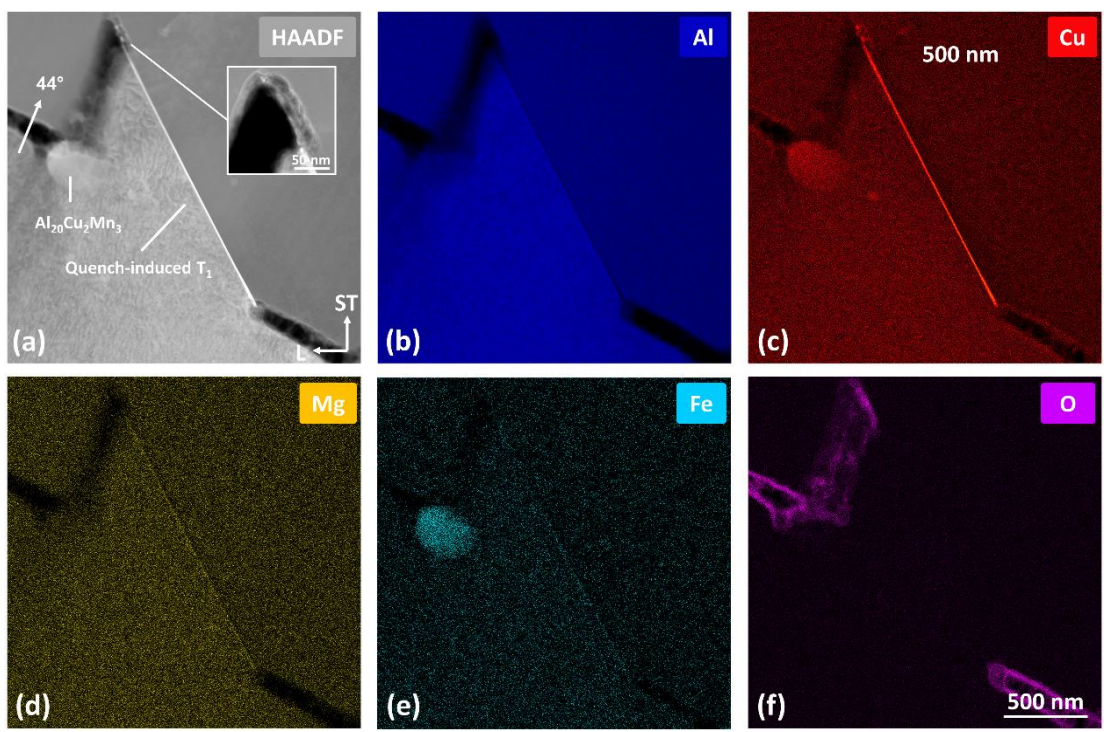

**Figure 11. (a) A HAADF-STEM micrograph and (b-f) the correlative STEM-EDX chemical maps showing details of intergranular corrosion and the associated $Al_{20}Cu_2Mn_3$ and quench-induced $T_1$ phases after 30 minutes of immersion. The number in (a) indicates the misorientation of grain boundary. The inset in (a) reveals details from the corroded region of $T_1$ precipitate. The location where the analysis was performed is indicated in Figure S6.**

In this region, intergranular corrosion propagates along a high angle grain boundary (i.e. ~ 44° in misorientation) and the particle-matrix interface of a $Al_{20}Cu_2Mn_3$ phase, Figure 11a. It also leads to the corrosion of a large quench-induced $T_1$ precipitate, with the associated grain boundaries corroded at the both ends. STEM-EDX analysis further reveals the Cu, O-rich remnants in the corroded regions with the relative depletion of Al, Mg and Fe, Figures 11b – 11e. Corrosion morphology was also observed in the matrix adjacent to the Cu, O-rich remnants. In addition, the relatively smaller particles of intergranular $T_1$ precipitates (~ 200 nm in length) on the same grain boundary were completely corroded to Cu, O-rich remnants within the crevice of intergranular corrosion, Figure S7.

Figure 12 further displays a HAADF-STEM micrograph and the correlative STEM-EDX chemical maps showing details of intergranular corrosion and the associated intermetallic phases in the region as indicated in Figure S6.

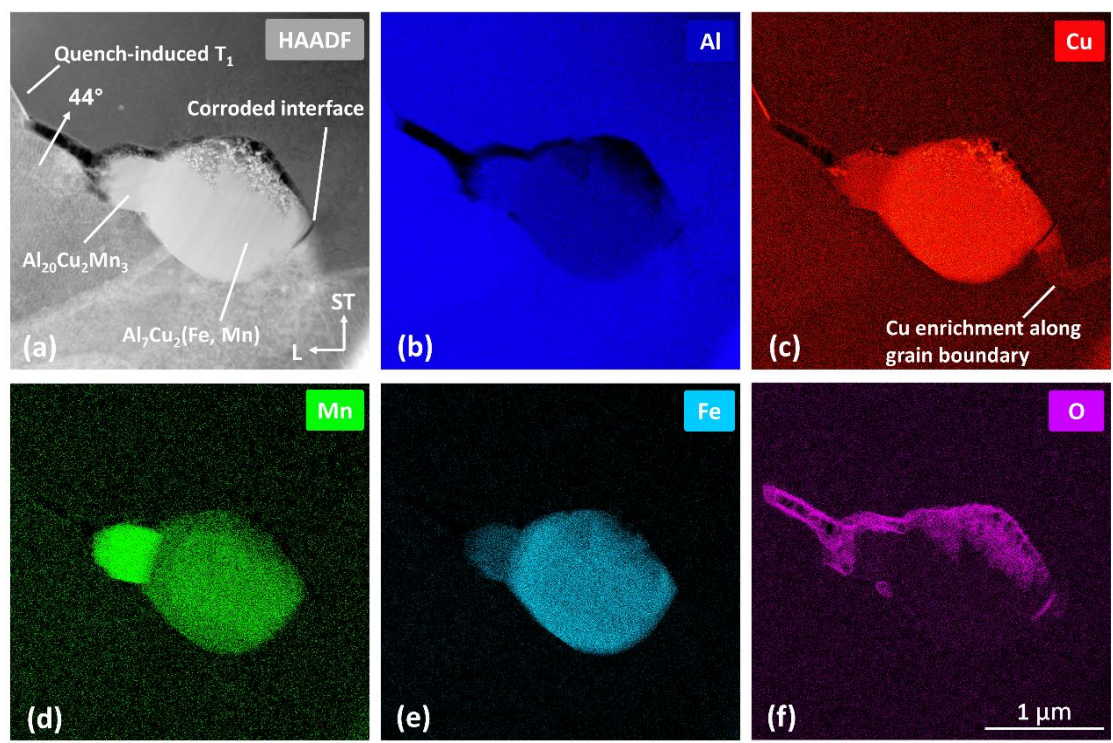

**Figure 12. (a) A HAADF-STEM micrograph and (b-f) the correlative STEM-EDX chemical maps showing details of the remnants of $Al_7Cu_2$(Fe, Mn) and $Al_{20}Cu_2Mn_3$ phases in the subsurface propagation area of intergranular corrosion after 30 minutes of immersion. The number in (a) indicates the misorientation of grain boundary. The location where the analysis was performed is indicated in Figure S6.**

In this region, the $Al_7Cu_2$(Fe, Mn) phase has the nanoporous morphology close to the corroded particle-matrix interface, whilst the $Al_{20}Cu_2Mn_3$ phase is not significantly changed. The corrosion of $Al_7Cu_2$(Fe, Mn) phase leads to the selective depletion of Al, Fe and Mn with the enrichment of Cu, Figures 12b – 12e. In particular, the decohesion of interface arising from corrosion attack was observed at the front end of corrosion crevice where the corrosion morphology of $Al_7Cu_2$(Fe, Mn) phase is not present. The enrichment of Cu within the regions ~ 170 nm in width was also observed along the associated grain boundary ahead of the corroded particle-matrix interface, Figure 12c. HR-TEM imaging further reveals the structural change to the lattice of $Al_7Cu_2$(Fe, Mn) phase along the <011> crystal direction and the corresponding reflections to Cu in the adjacent Al matrix close to the corroded particle-matrix interface, Figure S8.

## 4. Discussion

### 4.1 Initiation of intergranular corrosion

The examinations on top surface after 7 and 30 minutes of immersion clearly revealed initiation of intergranular corrosion from the clusters of constituent intermetallic phases associated with a dense network of grain boundaries. These clusters containing the noble $Al_7Cu_2$(Fe, Mn) and $Al_{20}Cu_2Mn_3$ phases act as effective cathodes that support oxygen reduction in the local region at top surface upon an exposure to electrolyte [25,37–39]. The cathodic reaction involving the formation of hydroxyl ions facilitates a local pH increase towards an alkaline environment, in which the protective oxide layer on the surrounding Al matrix is undermined [13,24,38]. This further leads to the formation of trenches adjacent to the cathodic particles via the anodic dissolution of the surrounding matrix [13,40].

In the present case, trenches with a maximum depth of ~ 2 μm were observed adjacent to the $Al_7Cu_2$(Fe, Mn) phases in the initiation site of intergranular corrosion after 7 minutes of immersion, whilst no trenches were observed adjacent to the $Al_{20}Cu_2Mn_3$ phases, Figure 3. Based on the existing studies on synthesized intermetallic compounds, corrosion potentials are not significantly varied between $Al_7Cu_2Fe$ and $Al_{20}Cu_2Mn_3$ in a neutral NaCl solution, whilst $Al_7Cu_2Fe$ is more electrochemically active and provides a higher cathodic current density than $Al_{20}Cu_2Mn_3$ [25,41]. In this case, the $Al_7Cu_2$(Fe, Mn) phases in the Al-Li alloys contain a higher level of Li than the $Al_{20}Cu_2Mn_3$ phases as indicated in Figure 1. This is consistent with observations from the existing studies conducted on the AA2050 and AA2099 alloys [17,43]. A higher level of Li in $Al_7Cu_2$(Fe, Mn) phases may further increase their electrochemical activity [17] and, in turn, promotes the initiation of intergranular corrosion in the early stage of corrosion immersion testing. In addition, the clusters containing large particles of $Al_7Cu_2(Fe, Mn)$ phase with a close interspacing distance facilitate a large cathode/anode area ratio that effectively supports the formation of trenches via anodic dissolution in the adjacent Al matrix [42], as observed in Figure 3.

The $Al_7Cu_2(Fe, Mn)$ phases containing Mg, Si and Zn at a trace level were corroded at top surface and close to the subsurface particle-matrix interface, Figures 3 and 4. This indicates that the $Al_7Cu_2(Fe, Mn)$ phases were electrochemically active and susceptible to corrosion attack in an alkaline environment established in the cathodic region. The corrosion of $Al_7Cu_2(Fe, Mn)$ at top surface involves the dissolution of Al, Mg, Si and Mn with the formation of a top layer where $Cu_2O$ and Cu coexist, Figures 4 and 5. The morphological characteristics and elemental distribution in the corroded region at top surface are similar to the de-alloyed intermetallic phases after immersion in an aerated NaCl electrolyte as previously reported in [27]. In this case, the de-alloying of $Al_7Cu_2(Fe, Mn)$ phase takes place via selective dissolution of less noble elements as driven by galvanic interaction between the relatively noble Cu-rich remnant and the intact part of intermetallic [44]. The selective dissolution of Al, Mg, Si, Mn and Fe was previously reported for synthesized compounds or intermetallic phases in acidic and neutral electrolytes [29,43,45], whilst Cu and Zn are relatively stable during de-alloying [27,46]. The formation of $Cu_2O$ in the top layer is mainly due to the oxidation of detached Cu debris released from the de-alloyed porous structure, while the oxidation of Cu to $Cu_2O$ may also occur at top surface after intermetallic particle was completely isolated from the surrounding matrix [27]. The de-alloying of $Al_7Cu_2(Fe, Mn)$ phase close to the subsurface particle-matrix interface leads to the Cu-rich remnant with a nanoporous structure similar to the existing observations from the de-alloyed intermetallic phases [29,47]. The de-alloyed structure formed in the subsurface region is relatively coarser than top surface. This is possibly due to an acidified electrolyte within the crevices along particle-matrix interface that impedes the repassivation of de-alloying process and, in turn, promotes porosity evolution [48].

*4.2 Propagation of intergranular corrosion*

The propagation of corrosion attack in the subsurface region is supported by an acidified chemical environment arising from hydrolysis of dissolved metal ions close to the front of trenches [49,50]. The acidified electrolyte effectively impedes repassivation and supports continuous corrosion attack along the particle-matrix interface and the associated grain boundary as observed in Figure 3. Crystallographic corrosion was also initiated from the interface of $Al_7Cu_2(Fe, Mn)$ phases and propagated in the adjacent grains containing the highly active Li-rich phases including $δ'$ and $T_1$ precipitates as previously reported in [51,52]. In addition, the cathodic reaction in the acidified electrolyte typically involves the formation of $H_2$ gas that creates a dynamic convection current within crevices [19,20]. This leads to the expulsion of corrosion products through the network of corroded grain boundaries and particle-matrix interface to the top surface of specimen [20], as observed in Figure 3. The formation of ring-like feature as observed in the periphery of initiation site (Figure 2c) is also related with a gel-like film of aluminium hydroxide filled with $H_2$ gas as previously proposed in [17,19].

The corrosion propagation along particle-matrix interface leads to the formation of a thin layer containing Al and O arising from the hydrolysis of dissolved Al ions, Figure 4. Due to a fast propagation rate along the particle-matrix interface, the $Al_7Cu_2(Fe, Mn)$ phase was completely isolated from the surrounding matrix after 7 minutes of immersion. Once isolated, the $Al_7Cu_2(Fe, Mn)$ phases at the initiation site stopped acting as the external cathodes that support propagation of intergranular corrosion down the surface. These particles then underwent self-dissolution including Cu at an increased corrosion potential towards a porous cluster containing $Cu_2O$ [27]. Due to a lack of galvanic interaction between the isolated intermetallic phases and the surrounding matrix, corrosion products were not formed at top surface adjacent to the isolated particles, Figures 6.

The strong intensity of $^7Li^+$ signal from the particle-matrix interface (i.e. as shown in Figure 1) was provided by the Li-rich $T_1$ precipitates as previously indicated in [30] and Figure S9. These precipitates abundant in Li are electrochemically active and susceptible to anodic dissolution due to galvanic interaction with the surrounding matrix [13,53,54]. The dissolution of $T_1$ phases usually leads to the remnant of Cu-rich nanoparticles as previously reported in [54]. In the present case, remnants from the dissolved $T_1$ precipitates were not observed at the particle-matrix interface of $Al_7Cu_2(Fe, Mn)$ phases in the initiation site of intergranular corrosion. This may be attributed to the expulsion of nanoparticles with the convection current created by the $H_2$ gas flow formed in an acidified chemical environment within crevices [19,20]. Alternatively, the absence of $T_1$ remnants suggests that the presence of $Al_7Cu_2$(Fe, Mn) phases alone has been sufficient for supporting the propagation of intergranular corrosion along the particle-matrix interface via galvanic interaction.

The propagation of intergranular corrosion away from the initiation site follows the network of grain boundary and the interface of intermetallic phases in the subsurface region, Figure 7. De-alloying of the $Al_7Cu_2$(Fe, Mn) phases associated with corrosion attack leads to the dissolution of Li, Mg, Al, Fe and Mn, resulting in a porous remnant containing Cu and Cu-oxides, Figures 8 – 10. De-alloying of intermetallic phases contributes to the propagation of intergranular corrosion by enhancing the acidity within the local crevice environment via the hydrolysis of dissolved metal ions. Corrosion morphology corresponding to de-alloying was not observed for the $Al_{20}Cu_2Mn_3$ phases associated with intergranular corrosion. However, the self-dissolution of a $Al_{20}Cu_2Mn_3$ particle being isolated from the surrounding matrix was observed after 30 minutes of immersion, Figure 9. Importantly, the examination of a $Al_7Cu_2$(Fe, Mn) phase at corrosion front has highlighted the particle-matrix interface as the preferred path for corrosion propagation ahead of the de-alloyed region, Figure 12. This indicates that corrosion propagation was facilitated by the anodic dissolution of Al matrix adjacent to the cathodic intermetallic phases with an effect of galvanic coupling. The noble particles of intermetallic phase also act as the internal cathodes that support corrosion propagation along the associated grain boundaries.

The redistribution of dissolved Li has been observed within the crevices where Na was detected after corrosion immersion testing, Figure 8. The co-existence of Li and Na in corrosion products suggests that Li was effectively dissolved in the acidified electrolyte containing $Na^+$, as previously reported in [43,53]. It is considered that the dynamic convection currents created by $H_2$ gas flow leads to the redeposition of dissolved Li ion at the locations with favourable geometric characteristics. This further leads to the remnants where Na and Li coexist as the specimen was dried after corrosion immersion testing. In addition, the redeposition of Mn within crevices as shown in Figure 8 has been related with the dissolution of $Al_{20}Cu_2Mn_3$ and $Al_7Cu_2$(Fe, Mn) phases. This is supported by the observation of Mn enrichment on the internal surface of crevice close to the dissolved particles of intermetallic phases,

Figure 9. The Cu-rich nanolayer along crevices as shown in Figures 8 and 10 is consistent with observations from an existing study conducted on AA6005 alloy [55]. The Cu-rich nanolayer was formed by the selective dissolution of Al along the corroded grain boundary via a de-alloying mechanism, which, in turn, acted as the internal cathode that supports continuous propagation of intergranular corrosion through the network of grain boundary [55].

The propagation of intergranular corrosion has been related with the anodic dissolution of grain boundary $T_1$ precipitates via a de-alloying mechanism as previously proposed in [10,56,57]. In the present case, Cu-rich nanoparticles arising from the de-alloying of $T_1$ precipitates [54] were observed within the crevice of corroded grain boundary, Figure S7. In addition, a partially dissolved quench-induced $T_1$ precipitate was observed along the corroded grain boundary, forming Cu-rich nanoparticles in the corroded region, Figure 11. The Cu-rich remnants of dissolved $T_1$ precipitate are relatively noble and act as internal cathodes that support the dissolution of the surrounding Al matrix with micro-galvanic interaction [13,54]. This is evidenced by the observation of corrosion morphology in the matrix adjacent to the corroded part of $T_1$ precipitate, Figure 11. In addition, corrosion propagation followed along the associated grain boundary with the intact part of $T_1$ precipitate remained, Figure 11. This suggests a competing mechanism of corrosion propagation between $T_1$ precipitates and the grain boundaries being susceptible to corrosion attack. While the detailed mechanism of corrosion propagation along grain boundary warrants further studies, the anodic dissolution of Cu-depleted PFZs and the variation of grain stored energy between the adjacent grains are the critical factors related with continuous corrosion attack along grain boundary at an absence of intergranular $T_1$ precipitates [9,14–16].

## 5. Conclusions

- Initiation of intergranular corrosion involves the formation of trenches adjacent to closely-spaced particles of large $Al_7Cu_2(Fe, Mn)$ phases, followed by corrosion propagation along the particle-matrix interface and the associated network of grain boundary.
- The $Al_7Cu_2(Fe, Mn)$ phases with a higher level of Cu and Li are more electrochemically active and susceptible to self-dissolution via a de-alloying mechanism as compared to the $Al_{20}Cu_2Mn_3$ phases.

- Corrosion propagation along the particle-matrix interface was facilitated by anodic dissolution of the surrounding Al matrix driven by the micro-galvanic interaction with cathodic intermetallic particles. The rapid propagation along the particle-matrix interface leads to the isolation of intermetallic phases from the early stage of corrosion immersion testing.
- The $Al_7Cu_2(Fe, Mn)$ and $Al_{20}Cu_2Mn_3$ phases in the subsurface region act as the internal cathodes that support continuous propagation of intergranular corrosion along the particle-matrix interface via anodic dissolution of the surrounding Al matrix.
- The $Al_7Cu_2(Fe, Mn)$, $Al_{20}Cu_2Mn_3$ and $T_1$ phase particles were dissolved along attacked grain boundary during corrosion process. The dissolved metal ions were redeposited through the network of crevice.

Charact. 141 (2018) 163–168.

**Supporting information**

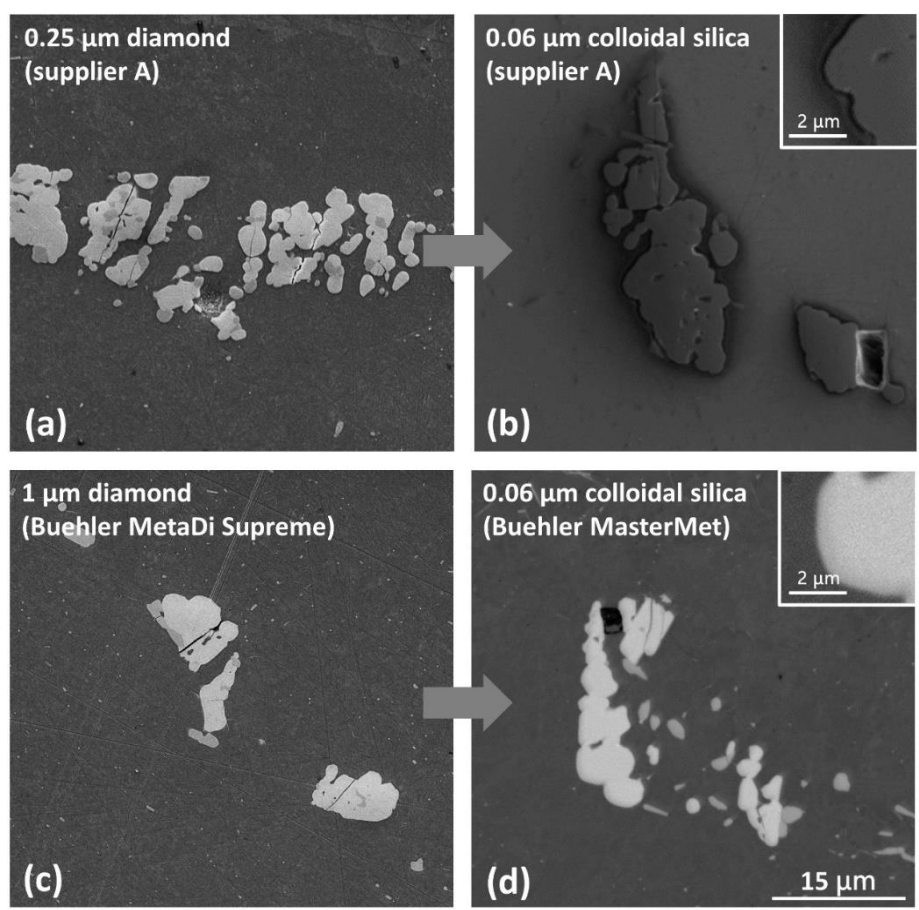

**Figure S1. BSE micrographs collected at 5 kV showing the surface of specimen after polishing with (a) 0.25 μm diamond suspension then (b) 0.06 μm colloidal silica suspension provided by Supplier A. Corrosion attack on intermetallic particles with trenching in the surrounding matrix were clearly observed. The same specimen was then grinded and prepared using (c) Buehler MetaDi Supreme 1 μm polycrystalline diamond suspension followed by (d) a solution of 1:1.5 Buehler Mastermet 0.06 μm colloidal silica suspension to deionised water. The corrosion attack to intermetallic particles was prevented. The insets in (b) and (d) further detail the surface topography at the interface of intermetallic particles with the surrounding matrix.**

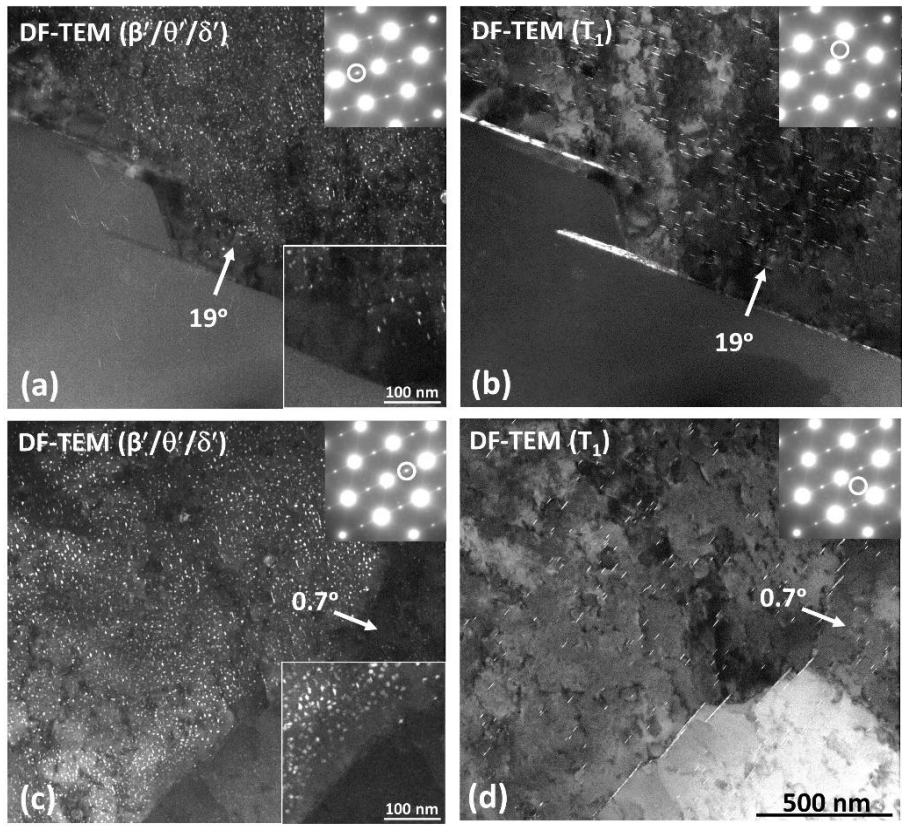

**Figure S2.** DF-TEM micrographs collected with the superlattice reflections of (a, c) the δ′/θ′/β′ precipitates and (b, d) the edge-on $T_1$ precipitates. The insets show the reflections selected for DF-TEM imaging and details of precipitates in a magnified view. The arrows indicate the misorientation of grain boundary.

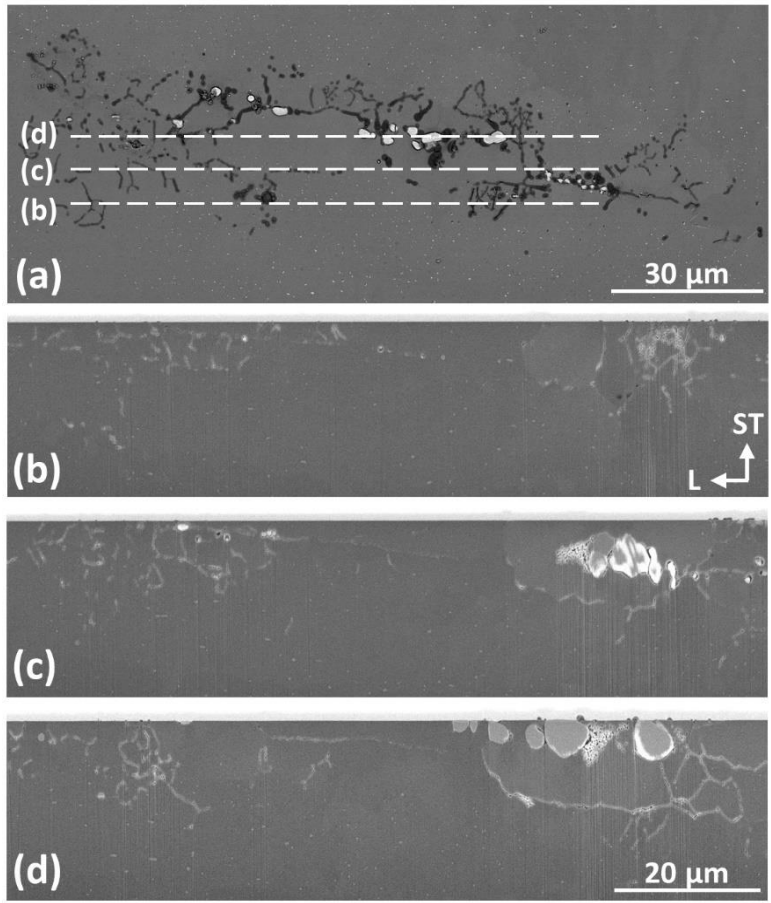

**Figure S3.** SE micrographs providing an overview of (a) surface and (b-d) cross-sections at the site of intergranular corrosion after 7 minutes of immersion. The cross-sections were fabricated using a FIB-SEM at the locations indicated by dash lines in (a).

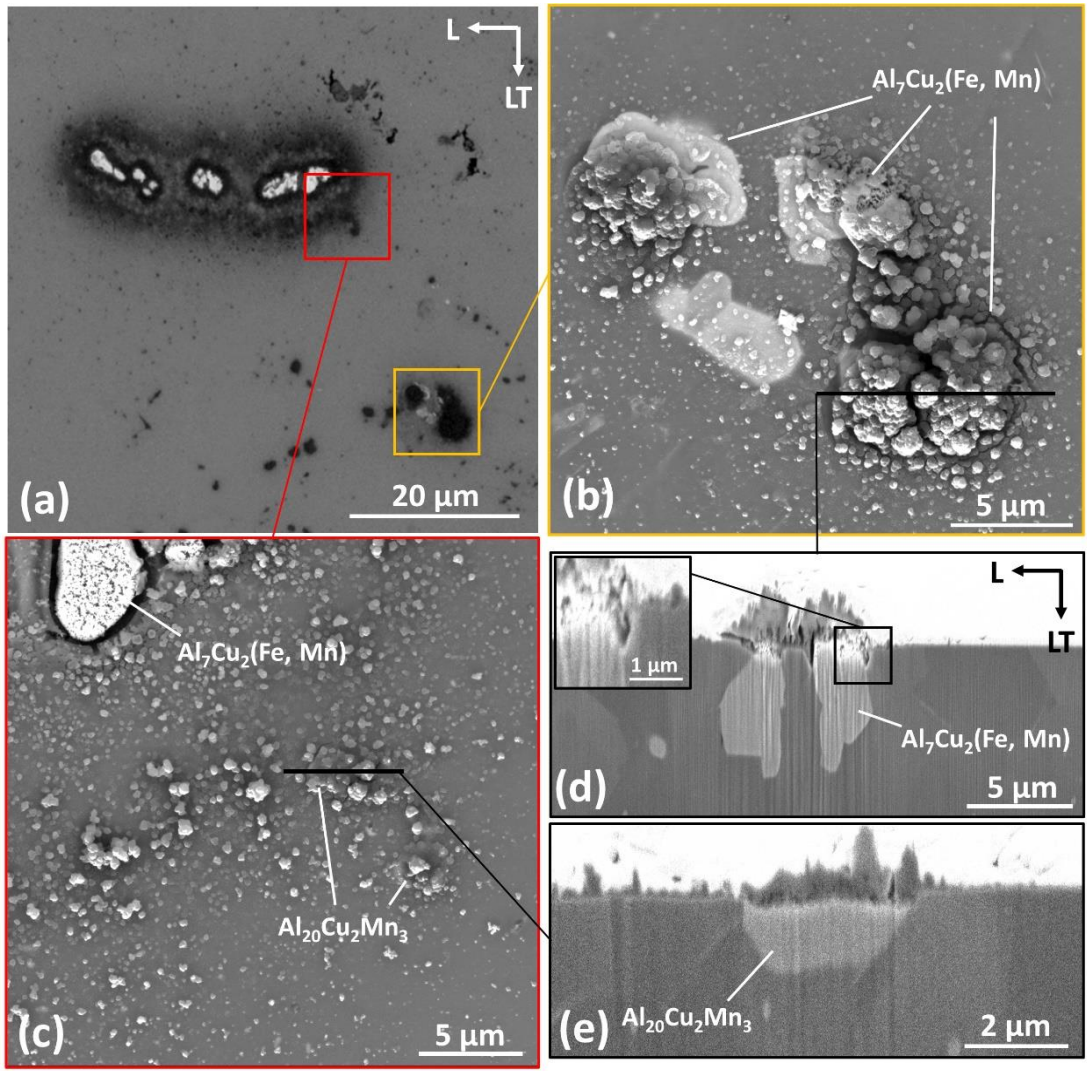

**Figure S4. (a – c)** BSE micrographs showing the intermetallic phases that are not related with intergranular corrosion after 30 minutes of immersion. **(d, e)** SE micrographs showing details of the targeted Al$_7$Cu$_2$(Fe, Mn) and Al$_{20}$Cu$_2$Mn$_3$ particles from cross-sections.

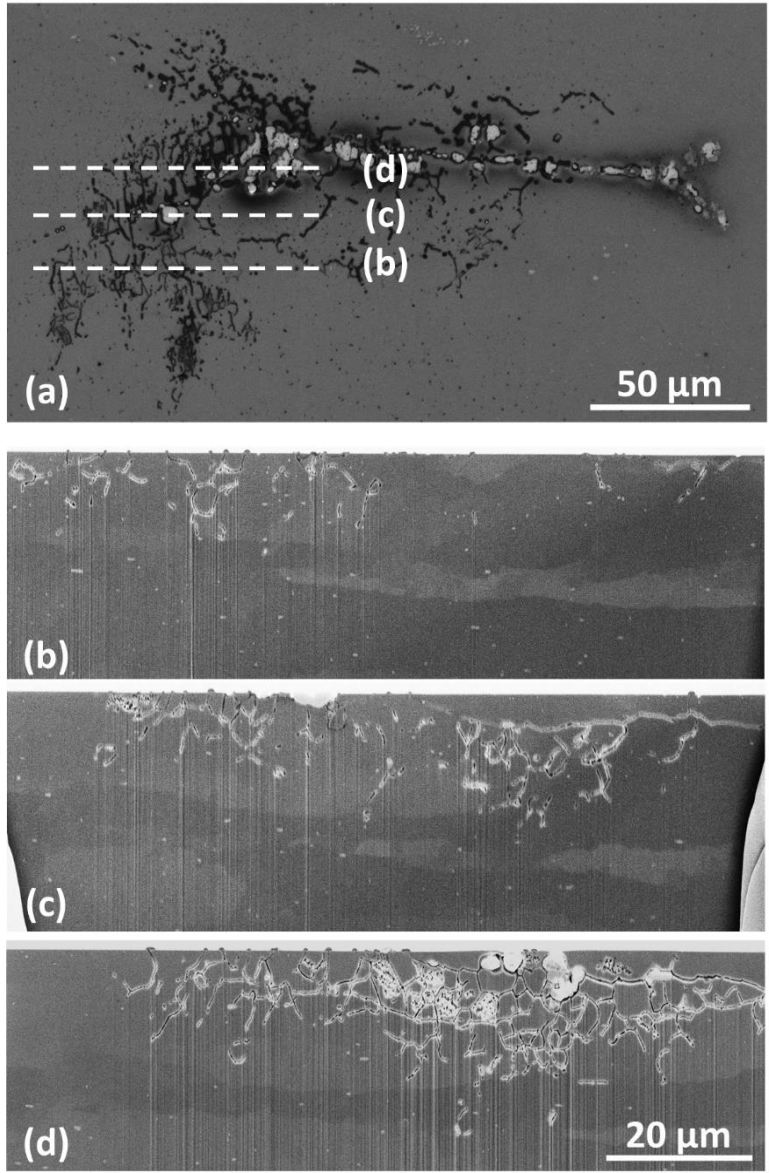

**Figure S5.** SE micrographs providing an overview of (a) surface and (b-d) cross-sections at the site of intergranular corrosion after 30 minutes of immersion. The cross-sections were fabricated using a FIB-SEM at the locations indicated by dash lines in (a).

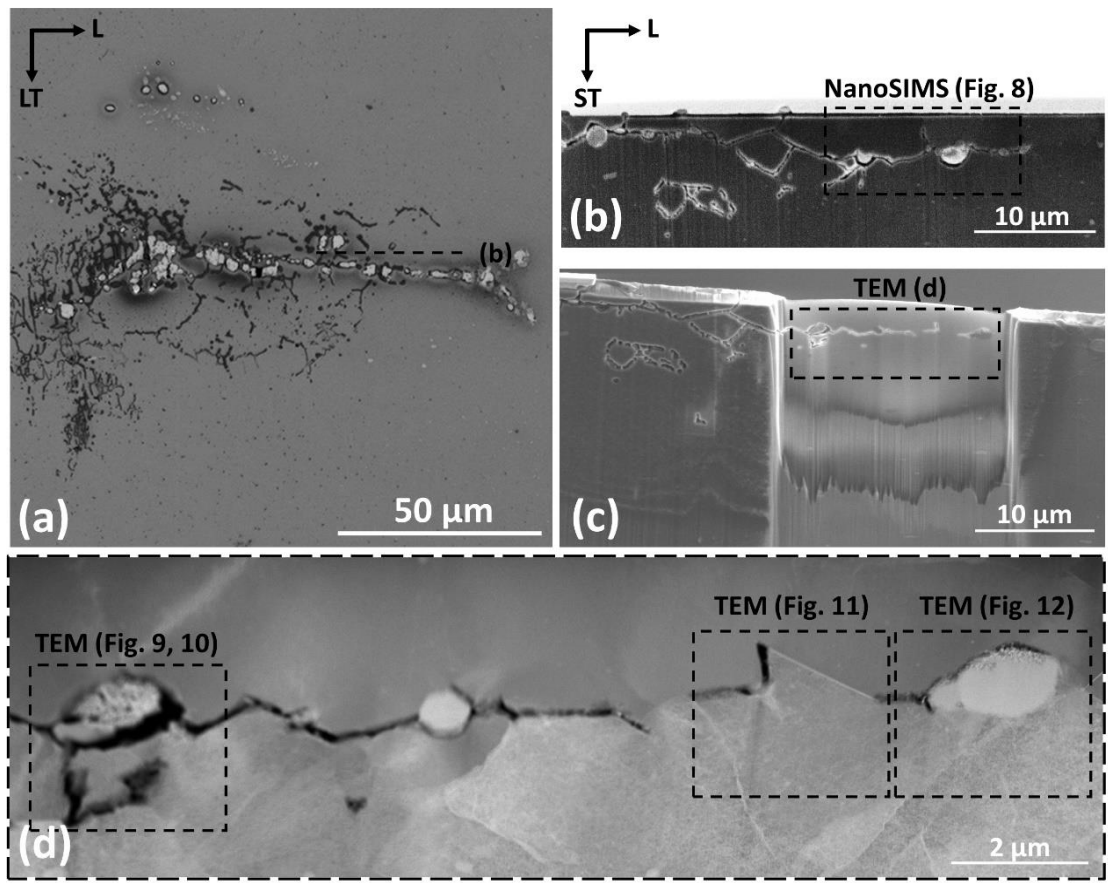

Figure S6. BSE micrograph showing (a) top-view and (b, c) cross-section of intergranular corrosion after 30 minutes of immersion. NanoSIMS and TEM analyses were conducted in the boxed areas in (b) and (c). (d) A HAADF-STEM micrograph showing an overview of TEM specimen. The detailed examinations as illustrated in Figures 9 – 12 were conducted in the boxed areas in (d).

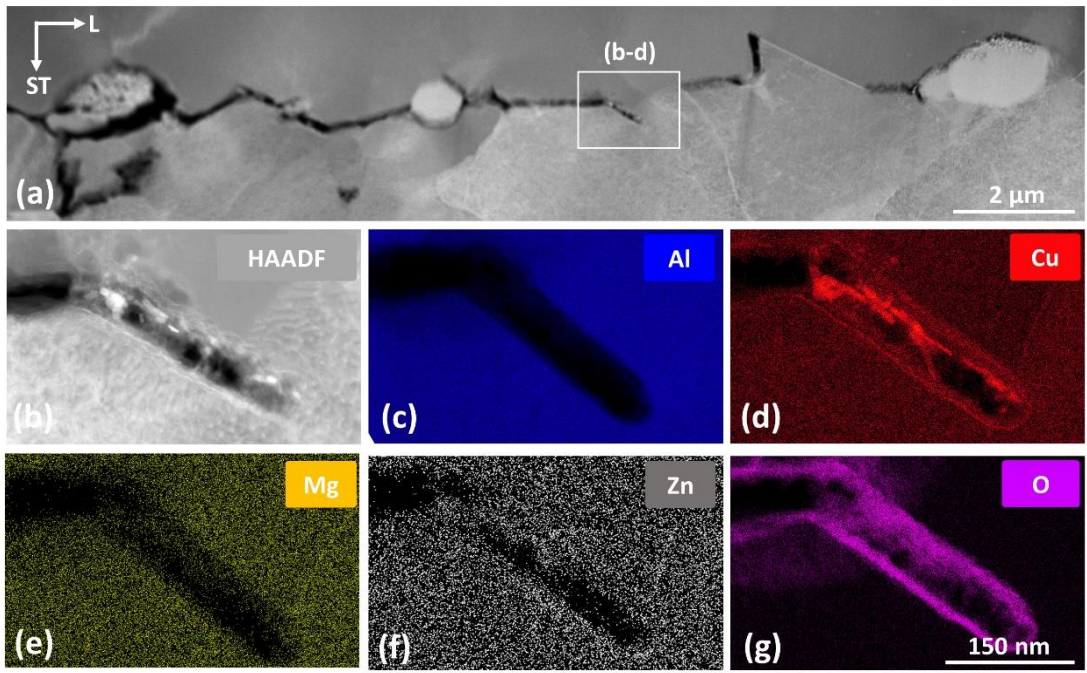

**Figure S7. HAADF-STEM micrographs showing (a) an overview and (b) the details of a corroded intergranular T$_1$ precipitate after 30 minutes of immersion. (c, d) STEM-EDX maps showing the distribution of Cu and O in the same region.**

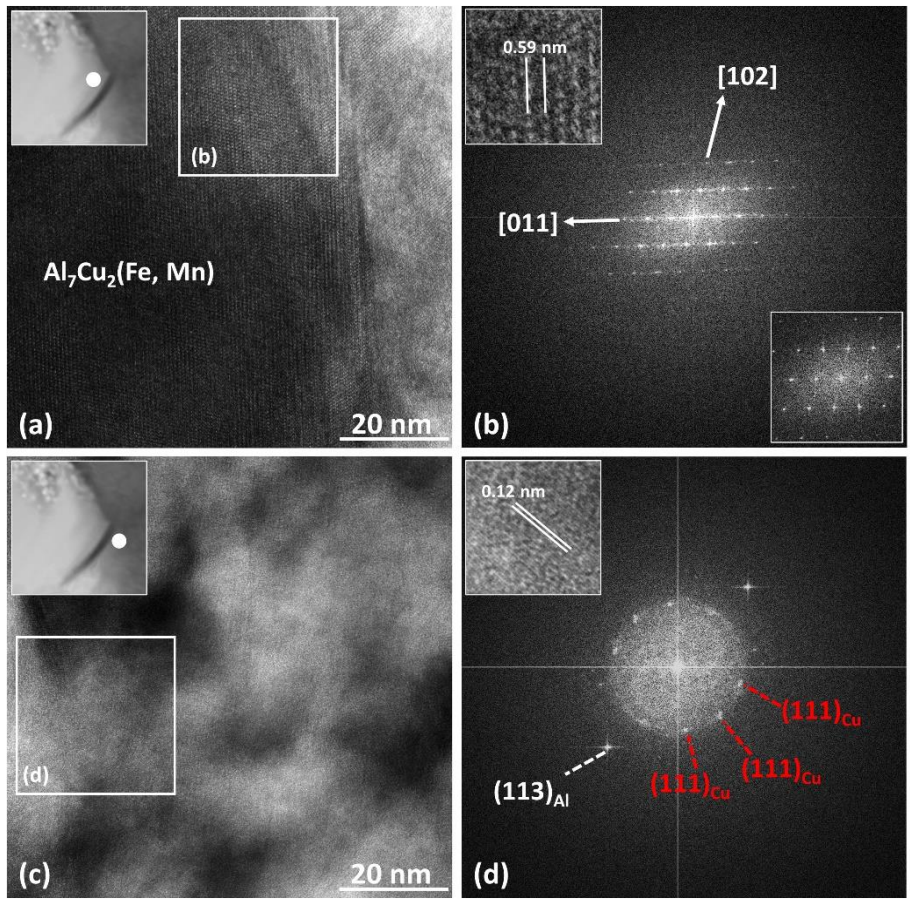

**Figure S8.** HR-TEM micrographs collected from (a) the Al₇Cu₂(Fe, Mn) phase and (c) the adjacent matrix close to the particle-matrix interface with (b, d) FFT patterns from the boxed regions. The insets in (a) and (c) indicate the locations where the HR-TEM micrographs were collected. The insets in (b) and (d) show details in the targeted regions and an FFT pattern from the intact part of Al₇Cu₂(Fe, Mn) phase.

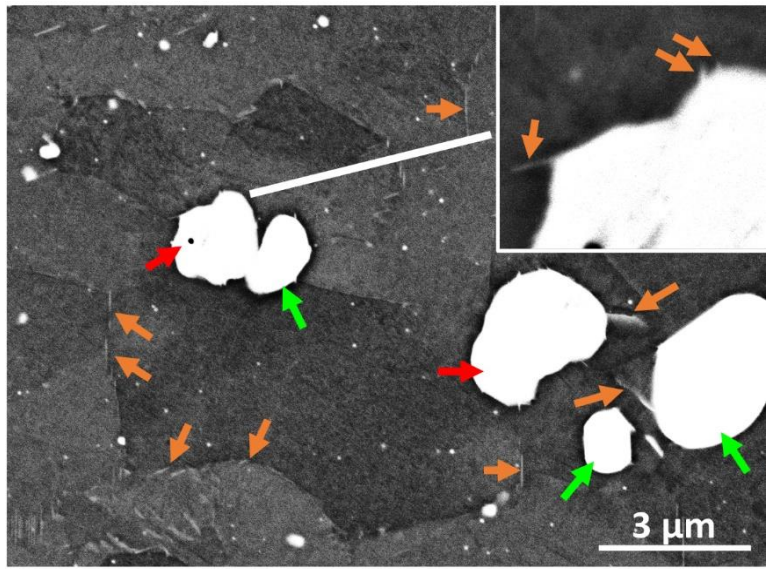

**Figure S9. A BSE micrograph showing $T_1$ precipitate formed close to the interface of $Al_7Cu_2$(Fe, Mn) and $Al_{20}Cu_2Mn_3$ phases with the surrounding matrix. The inset details the interfacial $T_1$ precipitates at a higher magnification.**